\documentclass[preprint,showpacs,preprintnumbers,amsmath,amssymb]{revtex4}

\usepackage{graphicx}
\usepackage{dcolumn}
\usepackage{bm}

\begin{document}

\title{Two Dimensional Nonlinear Nonequilibrium Kinetic Theory under Steady Heat Conduction}

\author{Kim Hyeon-Deuk \footnote{kim@kuchem.kyoto-u.ac.jp}}

\affiliation{Department of Chemistry, Kyoto
University, Kyoto 606-8502, Japan}

\date{\today}

\begin{abstract}
The two-dimensional steady-state Boltzmann equation for
 hard-disk molecules in the presence of a temperature gradient has been solved explicitly to second order in density and the temperature gradient. 
The two-dimensional equation of state and some physical quantities are calculated from it and compared with those
 for the two-dimensional steady-state Bhatnagar-Gross-Krook(BGK) equation and information theory. 
We have found that the same kind of qualitative differences as the three-dimensional case among these theories still appear in the two-dimensional case. 
\end{abstract}


\maketitle

\newpage

\section{Introduction}\label{history}
The behaviors of gases in nonequilibrium states have received
considerable attention from the standpoint of understanding the
characteristics of nonequilibrium
phenomena.\cite{chapman,resi,kogan,cercignani1,hand,sonebook,kaper,leb,cohen} 
The kinetic theory has contributed not only to the understanding of nonequilibrium
transport phenomena in gases but also to the development of general nonequilibrium statistical physics. 
It is well accepted that the Boltzmann equation is one of the most reliable
kinetic models for describing nonequilibrium phenomena in gas phases.
In the early stage of studies on the kinetic theory, 
great effort has been paid for solving kinetic model equations such as
the Boltzmann equation and deriving nonequilibrium velocity distribution
functions and
macroscopic nonequilibrium transport equations in terms of microscopic
molecular quantities. 
These attempts were strongly related to the development of
general nonequilibrium statistical physics such as linear nonequilibrium thermodynamics, Onsager's reciprocal theorem
 and the linear response theory.\cite{pomeau,onsager} 

Among various methods which give normal solutions of the Boltzmann
equation, the Chapman-Enskog method has been widely accepted as the most reliable method. 
It had been believed that Burnett determined the complete second-order solution of the Boltzmann
equation by the Chapman-Enskog method.\cite{chapman,burnett,gallis}
Physical importance of the second-order coefficients has been also demonstrated for
descriptions of shock wave profiles and sound
propagation phenomena.\cite{shocksound,shock1,foch} 
However, it was reported that Burnett's solution is not
 complete, and Schamberg derived the
precise velocity distribution function of the Boltzmann
equation to second order for Maxwell molecules.\cite{maxwell} 
On the other hand, because of its mathematical difficulty, the complete second-order solution of the Boltzmann equation for hard-core molecules has been derived quite recently.\cite{kim}  
Its validity has been also demonstrated by numerical experiments of both a molecular dynamics simulation and a direct simulation monte carlo method.\cite{fushiki,kimdoctor} 
Other kinetic models like the Bhatnagar-Gross-Krook
(BGK) equation\cite{bgk,lebo,korea,korea1,santos,santos1,santos4,santos2} have been
proposed mainly to avoid the mathematical difficulties in dealing with the
collision term of the Boltzmann equation.   
Although it was believed that its results approximately
agreed with those of the Boltzmann equation, some qualitative disagreements have been found in the second-order solutions.\cite{kim} 

Following its success and usefulness, the Boltzmann
equation is widely used in
order to describe various gas-phase transport phenomena such as granular
gases\cite{kim3,Noije,Espiov,Brey1},
plasma gases\cite{plasma1,plasma2}, polyatomic gases\cite{poly1}, relativistic gases\cite{relativ} and chemically reacting gases.  
Chemical reactions in gas phases have been studied with the aid of gas collision
theory. 
For the differential cross-section of
chemical reaction, the line-of-centers model proposed by Present has been accepted as a standard model to describe the chemical
reaction in gases.\cite{present,present10,fort,present2,present3,sigma,fort1,fort3}
This model can be derived explicitly using a collision law of hard-core
molecules, and the diameter of hard-core
molecules is regarded as a distance between centers of monatomic
molecules at contact.\cite{present,present10,sigma}
In equilibrium states, experimental results including
the temperature dependency can be fitted by
the results from the line-of-centers model.\cite{present,present2,present3} 
Under nonequilibrium situations such as gases under a heat
flux or a shear flow, their pure nonequilibrium contributions to the rate of
chemical reaction have also attracted much attention.\cite{fort,fort1,fort3,eu1,kim1,eu,nettleton1,nettleton2}
Since nonequilibrium correction terms of the chemical reaction rate are quadratic functions of
 nonequilibrium fluxes, the explicit nonequilibrium velocity distribution function of the Boltzmann equation for
hard-core molecules to second order is needed to derive it based on the line-of-centers model.\cite{fort,fort1,eu1,kim1} 
The pure effect of a heat flux on the chemical reaction rate has been recently calculated using the second-order velocity distribution function of the Boltzmann equation for
hard-core molecules.\cite{kim1}
In the letter, a \textit{thermometer} to measure a relation between a kinetic temperature of gases under a heat flux and a temperature of a heat bath has been also proposed. 

It is one of the most significant subjects in modern statistical
physics to construct the nonlinear nonequilibrium statistical mechanics and thermodynamics for a strongly nonequilibrium state
beyond the local equilibrium state, called the local nonequilibrium state. 
Zubarev\cite{zubarev,zubarev1} has developed nonequilibrium statistical mechanics and obtained the general form of a nonequilibrium velocity distribution
function with the aid of the maximum entropy principle. 
It is expanded to first order under some
constraints to obtain the first-order nonequilibrium velocity distribution function.\cite{katz} 
Jou and his coworkers have derived the nonequilibrium velocity distribution
function to second order by expanding it to second order under some
constraints, which is called information
theory.\cite{jou,jou0}  
Information theory has attracted interest in the development of a general
framework for nonlinear nonequilibrium statistical mechanics which can describe
the local nonequilibrium state.
The nonequilibrium velocity distribution function 
 from information theory has been applied to nonequilibrium systems, and some
 predictions were made. 
For example, in dilute gas systems under nonequilibrium fluxes, 
an anisotropic pressure and a nonequilibrium temperature which is not
identical with the kinetic temperature have been predicted.\cite{jou1,jou4,jou2,jou3}
There are also several applications of information theory to chemically
reacting gases.\cite{fort,fort1} 
However, some qualitative differences between information theory and 
the Boltzmann equation have been recently reported,
and the
invalidity of information theory as universal nonlinear nonequilibrium statistical mechanics has been demonstrated.\cite{kim2} 

There have been no efforts to solve two-dimensional kinetic models to second order and discuss the two-dimensional second-order nonequilibrium phenomena, though two-dimensional transport phenomena have created great interest.\cite{kawasaki,pomeau1,sengers,sengers1,ernst,oppenheim,alder}
The main aims of this paper are to reconstruct all the results obtained in refs.\cite{kim,kim1,kim2} in the case of two dimensional, and to discuss properties of the two-dimensional nonlinear nonequilibrium phenomena which reflect the local nonequilibrium state. 
In Secs.\ref{solve} and \ref{f2}, 
we have derived the explicit velocity distribution function of the two-dimensional steady-state Boltzmann
equation for hard-disk molecules to second order by the Chapman-Enskog method. 
In order to achieve that, we have extended the method we developed in ref.\cite{kim} to the two-dimensional case. 
We also obtain the nonequilibrium velocity distribution functions to second order for the two-dimensional steady-state BGK equation and information theory in Secs.\ref{solvebgk}
and \ref{itsolve}, respectively. 
All the nonlinear nonequilibrium velocity distribution functions are graphically compared in Sec.\ref{compare}. 
Using the two-dimensional nonequilibrium velocity distribution functions to second order, we discuss differences among those theories appearing in the two-dimensional nonlinear nonequilibrium transport phenomena in Sec.\ref{application}. 
In Sec.\ref{contributionCR}, we explain how to calculate the effect of steady heat flux on the rate
of chemical reaction based on the line-of-centers model in the two-dimensional case, and apply the two-dimensional nonequilibrium velocity distribution functions to second order 
to calculate it. 
We have also investigated dimensional dependency appearing in the nonlinear nonequilibrium phenomena which reflect the local nonequilibrium state. 
Our discussion and conclusion are given in Sec.\ref{ktdiscussion}.

\section{The Chapman-Enskog Method for Solving the two-dimensional Steady-State Boltzmann Equation}\label{solve}

\begin{figure}[htbp]
\includegraphics[width=7.cm]{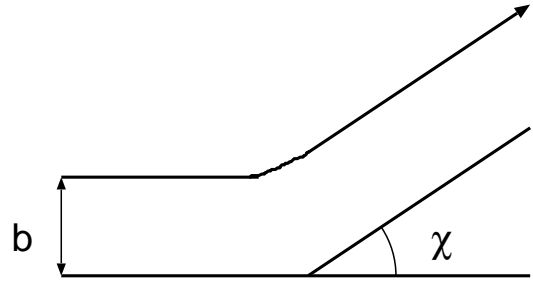} 
\caption{Schematic description of an interaction in two dimension.}
\label{interaction}
\end{figure}

Let us introduce the Chapman-Enskog method to solve the two-dimensional steady-state Boltzmann equation in this section. 
Assume that we have a system of dilute gases in a steady state, with velocity distribution
function $f_{\mathrm{1}}=f({\bf r},{\bf v}_{\mathrm{1}})$. 
The appropriate steady-state Boltzmann equation is 
\begin{equation}  
{\bf v}_{\mathrm{1}}\cdot\nabla f_{\mathrm{1}}=
J(f_{\mathrm{1}},f_{\mathrm{2}}),\label{be1}
\end{equation}
where the collision integral $J(f_{\mathrm{1}},f_{\mathrm{2}})$ is expressed as 
\begin{equation}  
J(f_{\mathrm{1}},f_{\mathrm{2}})\equiv
\int \int (f^{\prime}_{\mathrm{1}}f^{\prime}_{\mathrm{2}}-f_{\mathrm{1}}f_{\mathrm{2}})g {\mathrm d} b {\mathrm d}{\bf v}_{\mathrm{2}},\label{be2}
\end{equation}
with $f_{\mathrm{1}}^{\prime}=f({\bf r},{\bf v}_{\mathrm{1}}^{\prime})$
and $f_{\mathrm{2}}^{\prime}=f({\bf r},{\bf v}_{\mathrm{2}}^{\prime})$:  
${\bf v}_{\mathrm{1}}^{\prime}$ and ${\bf v}_{\mathrm{2}}^{\prime}$ are
postcollisional velocities of ${\bf v}_{\mathrm{1}}$ and ${\bf v}_{\mathrm{2}}$, respectively. 
The relative velocity of two molecules before and after an
interaction has the same magnitude $g=|{\bf v}_{\mathrm{1}}-{\bf
v}_{\mathrm{2}}|$
; the angle between the directions of the relative
velocity before and after the interaction is represented by $\chi$. 
The relative position of the two molecules is represented by $b$, called
the impact parameter. (see Fig.\ref{interaction})   
The impact parameter $b$ depends on kinds of the interaction between molecules, and one should specify the intermolecular interaction so as to
explicitly determine the impact parameter $b$ in the collision
term $J(f_{\mathrm{1}},f_{\mathrm{2}})$.  
Note that $\chi$ can be expressed as a function of $b$ for a central force. 

Suppose that the velocity distribution function $f_{\mathrm{1}}$ can be expanded as
\begin{equation}  
f_{\mathrm{1}}=f^{(0)}_{\mathrm{1}}+f^{(1)}_{\mathrm{1}}+f^{(2)}_{\mathrm{1}}+\cdots=f^{(0)}_{\mathrm{1}}(1+\phi^{(1)}_{\mathrm{1}}+\phi^{(2)}_{\mathrm{1}}+\cdots), \label{be3}
\end{equation}
where the small expansion parameter
will turn out the Knudsen number $K=l/L$, which means that the mean
free path of molecules $l$ should be much less than
the characteristic length $L$ for changes in macroscopic
variables. 
$f^{(0)}_{\mathrm{1}}$ is the local Maxwellian distribution function, written as 
\begin{equation}  
f^{(0)}_{\mathrm{1}}=\frac{n({\bf r}) m}{2\pi \kappa T({\bf r})}\exp\left[-\frac{m{\bf v}_{\mathrm{1}}^{2}}{2\kappa T({\bf r})}\right],\label{be3.5} 
\end{equation} 
with $m$ mass of the molecules and $\kappa$ the Boltzmann constant. 
$n({\bf r})$ and $T({\bf r})$ will be identified later as the density and the temperature at position ${\bf r}$, respectively. 
Substituting eq.(\ref{be3}) into the two-dimensional steady-state Boltzmann equation (\ref{be1}), we
arrive at the following set of equations which we will solve completely
in this paper:  
\begin{eqnarray}  
L[f^{(0)}_{\mathrm{1}}]\phi^{(1)}_{\mathrm{1}}={\bf v}_{\mathrm{1}}\cdot\nabla f^{(0)}_{\mathrm{1}}, \label{be4}
\end{eqnarray}
to first order and
\begin{eqnarray}  
L[f^{(0)}_{\mathrm{1}}]\phi^{(2)}_{\mathrm{1}}={\bf v}_{\mathrm{1}}\cdot\nabla f^{(1)}_{\mathrm{1}}-J(f^{(1)}_{\mathrm{1}},f^{(1)}_{\mathrm{2}}),\label{be5}
\end{eqnarray}
to second order. 
The linear integral operator $L[f^{(0)}_{\mathrm{1}}]$ is defined as
\begin{eqnarray}  
L[f^{(0)}_{\mathrm{1}}]X_{\mathrm{1}}\equiv \int \int f^{(0)}_{\mathrm{1}}f_{\mathrm{2}}^{(0)}(X_{\mathrm{1}}^{\prime}-X_{\mathrm{1}}+X^{\prime}_{\mathrm{2}}-X_{\mathrm{2}})g {\mathrm d}b {\mathrm d}{\bf v}_{\mathrm{2}}. \label{be6}
\end{eqnarray}
The solubility conditions of the integral equation (\ref{be4}) are given by
\begin{eqnarray}  
\int \Phi_{\mathrm{i}} {\bf v}_{\mathrm{1}}\cdot\nabla f^{(0)}_{\mathrm{1}} {\mathrm d}{\bf v}_{\mathrm{1}}=0,\label{be7}
\end{eqnarray}
where $\Phi_{\mathrm{i}}$ are collisional invariants:
\begin{eqnarray}  
\Phi_{1}=1, \quad \Phi_{2}=m{\bf v}_{\mathrm{1}}, \quad \Phi_{3}=\frac{1}{2}m{\bf v}_{\mathrm{1}}^{2}.\label{be8}
\end{eqnarray}
Substituting eq.(\ref{be3.5}) into the solubility conditions (\ref{be7}),
it is seen that $n\kappa T$ is uniform in the steady state. 
We use this result in our calculation to second order. 
Similarly, the solubility conditions of the integral equation (\ref{be5}) are given by
\begin{eqnarray}  
\int \Phi_{\mathrm{i}} {\bf v}_{\mathrm{1}}\cdot\nabla f^{(1)}_{\mathrm{1}} {\mathrm d}{\bf v}_{\mathrm{1}}=0,
\label{be9}
\end{eqnarray}
which will be considered in Sec.\ref{solubility}. 

To construct solutions of the integral equations
(\ref{be4}) and (\ref{be5}) definite, four further conditions must be
specified; we identify the density:   
\begin{equation}  
n({\bf r})\equiv\int f_{\mathrm{1}} {\mathrm d}{\bf v}_{\mathrm{1}}=\int f^{(0)}_{\mathrm{1}} {\mathrm d}{\bf v}_{\mathrm{1}},\label{be10}
\end{equation} 
the temperature:  
\begin{equation}  
n({\bf r})\kappa T({\bf r})\equiv
\int \frac{m{\bf v}_{\mathrm{1}}^{2}}{2}f_{\mathrm{1}} {\mathrm d}{\bf v}_{\mathrm{1}}=\int \frac{m{\bf v}_{\mathrm{1}}^{2}}{2}f^{(0)}_{\mathrm{1}} {\mathrm d}{\bf v}_{\mathrm{1}}, \label{be11}
\end{equation}
and the mean flow: 
\begin{equation}  
{\bf C}_{0}\equiv\int m {\bf v}_{\mathrm{1}}f_{\mathrm{1}} {\mathrm d}{\bf v}_{\mathrm{1}}=\int m {\bf v}_{\mathrm{1}}f^{(0)}_{\mathrm{1}} {\mathrm d}{\bf v}_{\mathrm{1}}.\label{be12}
\end{equation} 
Here we assume that no mean flow, \textit{i.e.} ${\bf C}_{0}=0$, exists in
the system. 
The introduction of these conditions distinguishes the Chapman-Enskog
adopted here from the Hilbert method in which the conserved quantities
are also expanded.\cite{resi} 
We assert that conditions (\ref{be10}), (\ref{be11}) and (\ref{be12}) do not
affect all our results in this paper. 
It should be noted that, to solve the integral equations (\ref{be4})
and (\ref{be5}), we should
consider only the case in which the right-hand sides of eqs.(\ref{be4})
and (\ref{be5}) are not zero: if the right-hand sides of eqs.(\ref{be4})
and (\ref{be5}) are zero, the integral equations (\ref{be4})
and (\ref{be5}) become homogeneous equations which do not have any
particular solutions.\cite{resi} 

\section{A Method for Solving the Integral Equations}\label{f2}
\subsection{A general form of the velocity distribution function}
To solve the integral equations (\ref{be4}) and (\ref{be5}),
We assume a general form of the velocity
distribution function: 
\begin{eqnarray}  
f_{\mathrm{1}}&=&f^{(0)}_{\mathrm{1}}\left[\sum_{r=0}^{\infty} r! B_{0r} S^{r}_{0}({\bf c}_{\mathrm{1}}^{2})+\sum_{k=1}^{\infty}\left( \frac{m}{2\kappa T} \right)^{\frac{k}{2}}\sum_{r=0}^{\infty} r! Y_{kr}({\bf c}_{\mathrm{1}}) S^{r}_{k}({\bf c}_{\mathrm{1}}^{2})\right],
\label{be13}
\end{eqnarray}
with ${\bf c}_{\mathrm{1}}\equiv (m/2\kappa T)^{1/2}{\bf v}_{\mathrm{1}}$
 the scaled velocity. 
$S_{k}^{p}(X)$ is a Sonine
polynomial, defined by 
\begin{eqnarray}  
(1-\omega)^{-k-1}e^{-\frac{X\omega}{1-\omega}}=\sum_{p=0}^{\infty} \Gamma(p+k+1) S_{k}^{p}(X)\omega^p,   
\label{be13.5}
\end{eqnarray}
and 
\begin{eqnarray}  
Y_{kr}({\bf c}_{\mathrm{1}}) \equiv B_{kr}Y_{k}({\bf c}_{\mathrm{1}})+C_{kr} Z_{k}({\bf c}_{\mathrm{1}}), \label{be14}
\end{eqnarray}
where $B_{kr}$ and $C_{kr}$ are coefficients to be
determined. 
Introducing the polar coordinate representation for ${\bf
c}_{{\mathrm{1}}}$, \textit{i.e.} $c_{{\mathrm{1}}x}=c_{{\mathrm{1}}}\cos \theta$, $c_{{\mathrm{1}}y}=c_{{\mathrm{1}}}\sin \theta$,  
\begin{eqnarray}
Y_{k}({\bf c}_{\mathrm{1}})= \left( \frac{2\kappa T}{m} \right)^{\frac{k}{2}}c_{\mathrm{1}}^{k} \cos k \phi,\label{be15}
\end{eqnarray}
and
\begin{eqnarray}
Z_{k}({\bf c}_{\mathrm{1}})=\left(\frac{2\kappa T}{m} \right)^{\frac{k}{2}}c_{\mathrm{1}}^{k} \sin k \phi. \label{be17}
\end{eqnarray}

Assumption of the velocity distribution function of eq.(\ref{be13}) has some mathematical advantages in our
calculation.  
Firstly, it is sufficient to determine the coefficients
$B_{kr}$, because $C_{kr}$ can be always determined from
$B_{kr}$ by transformations of axes. 
Secondly, some important physical quantities are related to 
coefficients $B_{kr}$ and $C_{kr}$: 
\textit{e.g.} the density (\ref{be10}), the temperature (\ref{be11}) and
the zero mean flow
(\ref{be12}) with $f_{\mathrm{1}}$ in
eq.(\ref{be13}) lead to the four equivalent
conditions: 
\begin{eqnarray}  
B_{00}=1,\quad B_{10}=C_{10}=0, \quad B_{01}=0. \label{be18}
\end{eqnarray}
Similarly, the pressure tensor $P_{ij}$ defined by
\begin{eqnarray}  
P_{ij}=\left(\frac{2\kappa T}{m}\right)^{2}\int^{\infty}_{-\infty}{\mathrm d}{\bf c}_{\mathrm{1}}mc_{{\mathrm{1}} i}c_{{\mathrm{1}} j}f_{\mathrm{1}},\label{be17.5}
\end{eqnarray}
for $i, j=x$ and $y$ is related to $B_{20}$ and $C_{20}$. 

The coefficients $B_{kr}$ except for those
in eq.(\ref{be18}) can be calculated as follows. 
Multiplying the two-dimensional steady-state Boltzmann equation (\ref{be1}) by 
\begin{eqnarray}  
Q_{kr}({\bf c}_{\mathrm{1}})\equiv 4\left(\frac{m}{2\kappa T}\right)^{\frac{k}{2}}\frac{Y_{k}({\bf c}_{\mathrm{1}})S^{r}_{k}({\bf c}_{\mathrm{1}}^{2})}{\Gamma(k+r+1)},\label{be20}
\end{eqnarray}
and then integrating over $(2\kappa T/m)^{1/2}{\bf c}_{\mathrm{1}}$, it is found that 
\begin{eqnarray}  
-\left(\frac{2\kappa T}{m}\right)^{\frac{1}{2}}<{\bf c}_{\mathrm{1}}\cdot\nabla Q_{kr}>_{av}+\nabla\cdot \left[\left(\frac{2\kappa T}{m}\right)^{\frac{1}{2}}<{\bf c}_{\mathrm{1}} Q_{kr}>_{av}\right]=\nonumber \\
\left(\frac{2\kappa T}{m}\right)^{2}\int \int \int (Q^{\prime}_{kr}-Q_{kr}) f_{\mathrm{1}}f_{\mathrm{2}} g {\mathrm d}b {\mathrm d}{\bf c}_{\mathrm{2}}{\mathrm d}{\bf c}_{\mathrm{1}},
\label{be19}
\end{eqnarray}
where $\Gamma(X)$ is the Gamma function, $<X>_{av}=(2\kappa T/m)\int X f_{\mathrm{1}} {\mathrm d}{\bf
c}_{\mathrm{1}}$ and $Q^{\prime}_{kr}$ represents the postcollisional
$Q_{kr}$. 
We should calculate both sides of eq.(\ref{be19}) for every $k$ and
$r$, because eq.(\ref{be19}) leads to simultaneous equations to determine $B_{kr}$.

For convenience, we introduce $\Omega_{kr}$ and $\Delta_{kr}$ as the left-hand 
and the right-hand sides of
eq.(\ref{be19}), respectively:    
\begin{equation}  
\Omega_{kr}\equiv-\left(\frac{2\kappa T}{m}\right)^{\frac{1}{2}}<{\bf c}_{\mathrm{1}}\cdot\nabla Q_{kr}>_{av}+\nabla\cdot \left[\left(\frac{2\kappa T}{m}\right)^{\frac{1}{2}}<{\bf c}_{\mathrm{1}} Q_{kr}>_{av}\right],\label{be19.0}
\end{equation}
and
\begin{equation}  
\Delta_{kr}\equiv\left(\frac{2\kappa T}{m}\right)^{2}\int \int \int (Q^{\prime}_{kr}-Q_{kr}) f_{\mathrm{1}}f_{\mathrm{2}}g {\mathrm d}b {\mathrm d}{\bf c}_{\mathrm{2}}{\mathrm d}{\bf c}_{\mathrm{1}}.\label{be19.1}
\end{equation}
We will calculate $\Omega_{kr}$ and $\Delta_{kr}$ separately. 
From Appendix \ref{a3}, the result of $\Omega_{kr}$ becomes 
\begin{eqnarray}  
\Omega_{kr}&=& n \left(\frac{2\kappa T}{m}\right)^{\frac{1}{2}} \left[(r+\frac{k}{2}-\frac{1}{2})(D_{k,r}\frac{\partial_{x} T}{T}+E_{k,r}\frac{\partial_{y} T}{T})\right. \nonumber \\
& & \left. -(D_{k,r-1}\frac{\partial_{x} T}{T}+E_{k,r-1}\frac{\partial_{y} T}{T})+\partial_{x} D_{k,r}+\partial_{y} E_{k,r}\right], 
\label{be25.5}
\end{eqnarray}
where $\partial_{\mathrm{i}} X$ denotes $\partial X/\partial \mathrm{i}$ for $\mathrm{i}=x$ and $y$.  
$D_{k,r}$ and $E_{k,r}$ are functions of $B_{kr}$ and $C_{kr}$, as is written in Appendix \ref{a3}. 

\subsection{The collision term $\Delta_{\lowercase{kr}}$}\label{delta}

Next we calculate the collision term $\Delta_{kr}$ in eq.(\ref{be19.1}). 
We should specify the kind of the interaction of molecules so as to perform the calculation
of the collision
term $\Delta_{kr}$. 
For hard-disk molecules, the impact parameter $b$ is given by the relation
\begin{eqnarray}  
b=d \cos \frac{\chi}{2},\label{be26}
\end{eqnarray}
where $d$ is the hard-disk molecular diameter. 
The collision differential cross section is obtained by 
\begin{eqnarray}  
{\mathrm d} b=-\frac{d}{2} \sin\frac{\chi}{2} d\chi,\label{be260}
\end{eqnarray} 
Therefore, $\Delta_{kr}$ for hard-disk molecules, $\Delta_{kr}^{\mathrm{H}}$, becomes
\begin{eqnarray}
\Delta_{kr}^{\mathrm{H}}&=&\frac{d}{2}\int_{0}^{2 \pi} [F_{kr}^{1}(\chi)-F_{kr}^{1}(0) ]\sin\frac{\chi}{2} {\mathrm d}\chi, \label{be28}
\end{eqnarray}
where $F_{kr}^{\mu}(\chi)$ is defined as
\begin{eqnarray}
F_{kr}^{\mu}(\chi)\equiv\left(\frac{2\kappa T}{m}\right)^{2} \int \int Q^{\prime}_{kr} f_{\mathrm{1}}f_{\mathrm{2}}g^{\mu} {\mathrm d}{\bf c}_{\mathrm{2}}{\mathrm d}{\bf c}_{\mathrm{1}}, \label{be27}
\end{eqnarray}
and we have used $F_{kr}^{\mu}(0)=(2\kappa T/m)^{2}\int\int Q_{kr} f_{\mathrm{1}}f_{\mathrm{2}}g^{\mu} {\mathrm d}{\bf c}_{\mathrm{2}}{\mathrm d}{\bf c}_{\mathrm{1}}$. 
Note that $\chi=0$ if $b > d$.  
 
From eq.(\ref{be28}), it is sufficient to calculate $F_{kr}^{1}(\chi)$ for deriving $\Delta_{kr}^{\mathrm{H}}$. 
The details of $F_{kr}^{1}(\chi)$ are written in Appendix \ref{a2}. 
Several explicit forms of $\Delta_{kr}^{\mathrm{H}}$ are also
demonstrated in Appendices \ref{a10} and \ref{a50}. 
From the definitions (\ref{be19.0}) and (\ref{be19.1}), both sides of eq.(\ref{be19}) for arbitrary $k$ and
$r$ can be calculated for hard-disk molecules via 
\begin{eqnarray}
\Omega_{kr}^{\mathrm{H}}=\Delta_{kr}^{\mathrm{H}},\label{be0}
\end{eqnarray}
which produces a set of simultaneous equations determining the coefficients
$B_{kr}$, as is explained in Appendices \ref{a10} and \ref{a50}. 
Here $\Omega_{kr}^{\mathrm{H}}$ denotes $\Omega_{kr}$ for hard-disk
molecules. 

\subsection{Determination of $B_{kr}$}\label{coefficient}
We will determine the first-order coefficients
$B_{kr}^{\mathrm{I}}$ and the second-order coefficients
$B_{kr}^{\mathrm{II}}$ in accordance with the previous two subsections, which corresponds to solving the integral equations (\ref{be4})
and (\ref{be5}), respectively. 
Here the upper suffices $\mathrm{I}$ and $\mathrm{II}$ are introduced to
specify the order of $K$. 

\subsubsection{The First Order}\label{first}
We show the results of the first-order coefficients
 $B_{kr}^{\mathrm{I}}$ of which the
solution of the integral equation (\ref{be4}), $\phi^{(1)}_{\mathrm{1}}$, is composed. 
They can be written in the form: 
\begin{eqnarray}
B_{kr}^{\mathrm{I}}=\delta_{k,1} b_{1r}\frac{2\partial_{x} T}{\sqrt{2\pi}d n T}. \label{be37}
\end{eqnarray}
Values of the constants $b_{1r}$ are given in Table \ref{b1r}. 
The calculation of $B_{kr}^{\mathrm{I}}$ is explained in Appendix
\ref{a10}. 
It is seen that $B_{kr}^{\mathrm{I}}$ is of the order of the Knudsen
number $K$. 
Though $B_{kr}^{\mathrm{I}}$ was derived only to the lowest order
approximation\cite{sengers}, \textit{i.e.} $B_{kr}^{\mathrm{I}}$ for $r=1$, 
we have 
obtained $B_{kr}^{\mathrm{I}}$ for $r\le 7$ in this paper. 
This ensures the convergence of all the physical quantities which will
be calculated in this paper. 
It should be mentioned that our value of $B_{kr}^{\mathrm{I}}$ for
the lowest Sonine approximation, \textit{i.e.} $r=1$, is
identical with Sengers's value\cite{sengers}. 
Once $B_{kr}^{\mathrm{I}}$ have been calculated,
$C_{kr}^{\mathrm{I}}$ can be written down directly by replacing $\partial_{x} T$
by $\partial_{y} T$ by symmetry. 
Substituting all the first-order coefficients derived here into eq.(\ref{be14}),
we can finally obtain the first-order velocity distribution function $f^{(1)}_{\mathrm{1}}$.

\subsubsection{Solubility Conditions for $\phi^{(2)}_{\mathrm{1}}$}\label{solubility}
Since the first-order velocity distribution function $f^{(1)}_{\mathrm{1}}$ has been obtained, 
the solubility conditions for the integral equation (\ref{be5}) should be considered before we
attempt to derive the explicit second-order solution $\phi^{(2)}_{\mathrm{1}}$. 
The solubility conditions for $\phi^{(2)}_{\mathrm{1}}$, eqs.(\ref{be9}), lead to
the condition
\begin{eqnarray}  
\nabla \cdot {\bf J}^{(1)}=0, \label{be38}
\end{eqnarray}
where ${\bf J}^{(1)}$, \textit{i.e.} the heat flux for $f^{(1)}_{\mathrm{1}}$, can be obtained as  
\begin{eqnarray}  
{\bf J}^{(1)}&\equiv&\left(\frac{2\kappa T}{m}\right)^{\frac{5}{2}}\int^{\infty}_{-\infty}{\mathrm d}{\bf c}_{\mathrm{1}}\frac{m{\bf c}_{\mathrm{1}}^{2}}{2}{\bf c}_{\mathrm{1}} f^{(1)}_{\mathrm{1}} \label{be39A} \\
&=&-b_{11}\left(\frac{\kappa T}{\pi m}\right)^{\frac{1}{2}}\frac{2\kappa}{d}\nabla T, \label{be39}
\end{eqnarray} 
with the appropriate value for $b_{11}$ listed in Table \ref{b1r}. 
It must be emphasized that, since ${\bf J}^{(2)}$, \textit{i.e.} the heat flux for
$f^{(2)}_{\mathrm{1}}$, does not appear, the solubility conditions of
the two-dimensional steady-state Boltzmann equation for $\phi^{(2)}_{\mathrm{1}}$ lead to
the heat flux being constant to second order. 
This fact is in harmony with a general property that the total heat flux should be uniform in the steady state.
From eqs.(\ref{be38}) and (\ref{be39}), we also obtain an important relation
between $\left(\nabla T\right)^{2}$ and $\nabla^{2} T$,  
\begin{eqnarray}
\frac{\left(\nabla T\right)^{2}}{2T}+\nabla^{2} T=0.  
\label{be39.5}
\end{eqnarray} 
Owing to the relation (\ref{be39.5}), terms of $\nabla^{2} T$ can be replaced by terms of $\left(\nabla T\right)^{2}$. 

\subsubsection{The Second Order}\label{second}
We write down the results of the second-order coefficients $B_{kr}^{\mathrm{II}}$ of which $\phi^{(2)}_{\mathrm{1}}$ is composed. 
Using the relation (\ref{be39.5}), we can determine the second-order coefficients
 $B_{0r}^{\mathrm{II}}$ appearing in
eq.(\ref{be13}) as 
\begin{eqnarray}
B_{0r}^{\mathrm{II}}=\frac{b_{0r}}{\pi d^{2}n^{2}T^{2}}\left(\nabla T\right)^{2}. \label{be42AB}
\end{eqnarray}
Values for the constants $b_{0r}$ are summarized in Table \ref{b0r}. 
The calculation of $B_{0r}^{\mathrm{II}}$ is shown in Appendix
\ref{a50}. 
We have calculated $B_{0r}^{\mathrm{II}}$ to $7$th approximation,
\textit{i.e.} $B_{0r}^{\mathrm{II}}$ for $r\le 6$ in this paper. 

The other second-order coefficients
 $B_{kr}^{\mathrm{II}}$ in eq.(\ref{be14}) can be written in the
final form: 
\begin{eqnarray}
B_{kr}^{\mathrm{II}}=\frac{\delta_{k,2}}{\pi d^{2}n^{2}T^{2}}
\left\{b_{2r}^{A}\left[\left(\partial_{x} T\right)^{2}-\left(\partial_{y} T\right)^{2}\right]+b_{2r}^{B}T\left[\partial_{x}^{2} T-\partial_{y}^{2} T\right]\right\}. \label{be45AB}
\end{eqnarray}
Values for the constants $b_{2r}^{A}$ and $b_{2r}^{B}$ are summarized in Table \ref{b2rAB}. 
The calculation of $B_{kr}^{\mathrm{II}}$ is explained in Appendix
\ref{a50}. 
Owing to the properties of the
trigonometrical function, $C_{kr}^{\mathrm{II}}$ can be obtained by
replacing $(\partial_{x} T)^{2}-(\partial_{y} T)^{2}$ and $\partial_{x}^{2} T-\partial_{y}^{2} T$ in eq.(\ref{be45AB})
by $2 \partial_{x} T \partial_{y} T$ and
$2 T \partial_{x}\partial_{y} T$, respectively, using an axis change. 

One can see that both of $B_{0r}^{\mathrm{II}}$ and
$B_{kr}^{\mathrm{II}}$ are of the order of $K^2$. 
As is also found in three dimension case\cite{kim}, we have confirmed the fact that
$B_{kr}^{\mathrm{II}}$ for $k=4$ and $6$ do not appear. 
This fact strongly suggests that $B_{kr}^{\mathrm{II}}$ for all $k$ greater than $2$ do not appear, which was also expected in ref.\cite{kim} and recently confirmed in ref.\cite{fushiki}.  
We finally obtain
$f^{(2)}_{\mathrm{1}}$ by substituting the second-order coefficients
obtained here into eqs.(\ref{be13}) and (\ref{be14}).  
Finally, we note that, though we have derived all the constants
$b_{1r}$, $b_{0r}$, $b_{2r}^{A}$ and $b_{2r}^{B}$ in forms of fractions, we have written
them in forms of four significant figures in this paper, since the forms of those fractions are
too complicated. 

\subsubsection{The Velocity Distribution Function to Second Order}\label{f2x}
The velocity distribution function for hard-disk molecules which we
 have derived in this subsection valid to second
 order is now applied to a nonequilibrium steady-state system under the
 temperature gradient along $x$-axis. 
In this case, the form of $B_{0r}^{\mathrm{II}}$ in eq.(\ref{be42AB}) becomes
\begin{eqnarray}
B_{0r}^{\mathrm{II}}=\frac{b_{0r}}{\pi d^{2}n^{2}T^{2}}\left(\partial_{x} T\right)^{2}, \label{be42}
\end{eqnarray}
 and, using the relation (\ref{be39.5}), $B_{kr}^{\mathrm{II}}$ in eq.(\ref{be45AB}) can be
transformed into a more simple form: 
\begin{eqnarray}
B_{kr}^{\mathrm{II}}=\frac{\delta_{k,2}b_{2r}}{\pi d^{2}n^{2}T^{2}}\left(\partial_{x} T\right)^{2}, \label{be45}
\end{eqnarray}
where values for the constants $b_{2r}$ are summarized in
Table \ref{b2r}. 
The other second-order term $C_{kr}^{\mathrm{II}}$ becomes zero.

From eqs.(\ref{be13}) and (\ref{be14}), the velocity distribution function of the steady-state Boltzmann
equation for hard-disk molecules to second order in the
 temperature gradient along $x$-axis can be written as
\begin{eqnarray}
f&=&f^{(0)} \left \{ 1-\frac{J_{x}}{b_{11} n\kappa T}\left(\frac{m}{2\kappa T}\right)^{\frac{1}{2}}\sum_{r\ge 1}r! b_{1r}c_{x} S^{r}_{1}({\bf c}^{2}) \right. \nonumber\\
& &\left.+\frac{mJ_{x}^{2}}{4 b_{11}^{2}n^{2}\kappa^{3}T^{3}}\left[\sum_{r\ge 2}r! b_{0r}S^{r}_{0}({\bf c}^{2})+\sum_{r\ge 0}r! b_{2r}(c_{x}^{2}-c_{y}^{2})S^{r}_{2}({\bf c}^{2})\right] \right\}, 
\label{be46}
\end{eqnarray}
where the specific values for $b_{1r}$, $b_{0r}$ and $b_{2r}$ are found in
Tables \ref{b1r}, \ref{b0r} and \ref{b2r}, respectively.   
$J_{x}$ corresponds to the $x$ component of
 the heat flux in eq.(\ref{be39}). 
Note that we have changed ${\bf c}_{\mathrm{1}}$ to ${\bf c}$. 
As can be seen from eq.(\ref{be46}), 
the explicit form of the velocity distribution function for hard-disk molecules becomes
the sum of an infinite series of Sonine
polynomials. 

\begin{figure}[htbp]
\includegraphics[width=15.cm]{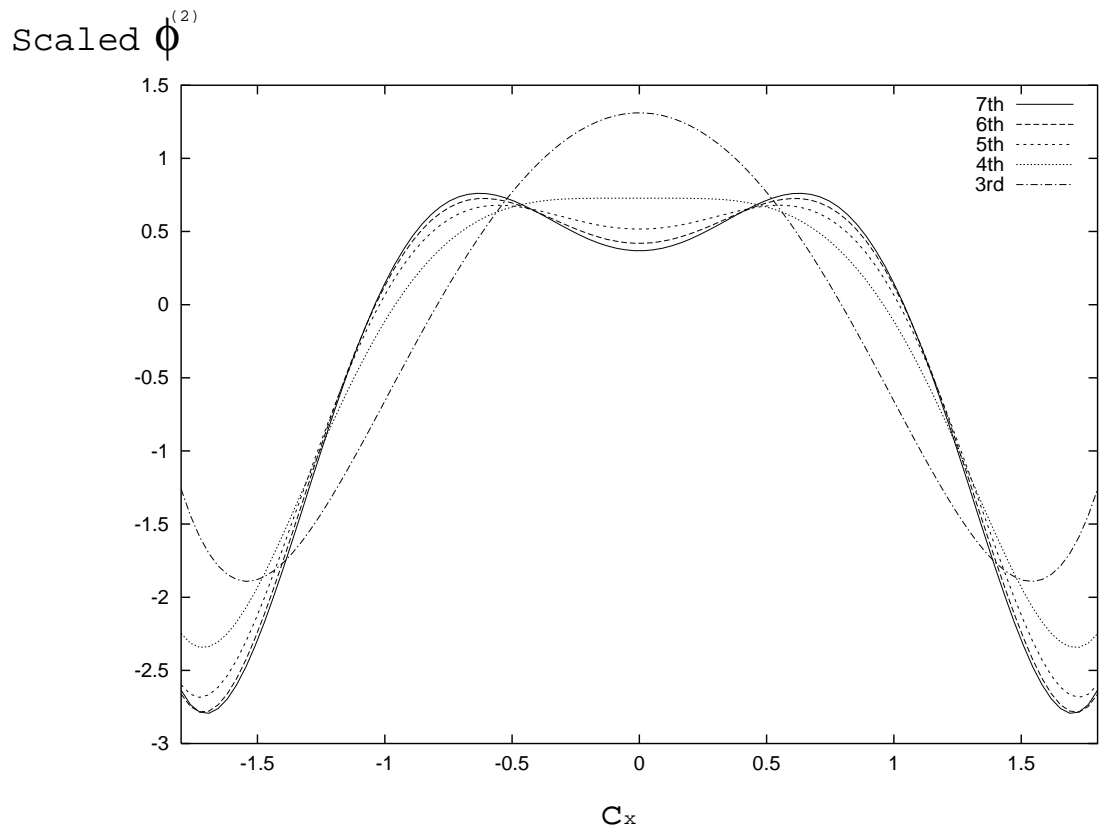} 
\caption{The scaled $\phi^{(2)}$s for hard-disk
 molecules. 
The dash-dotted line, the dotted line, the short-dashed line, the long-dashed line and the solid line correspond to the scaled $\phi^{(2)}$ for hard-disk
 molecules with the $3$th, $4$th, $5$th, $6$th and $7$th approximation $b_{0r}$ and
 $b_{2r}$, respectively. 
Note that we put $c_{y}=0$.}
\label{MDf2}
\end{figure}

\begin{figure}[htbp]
\includegraphics[width=15.cm]{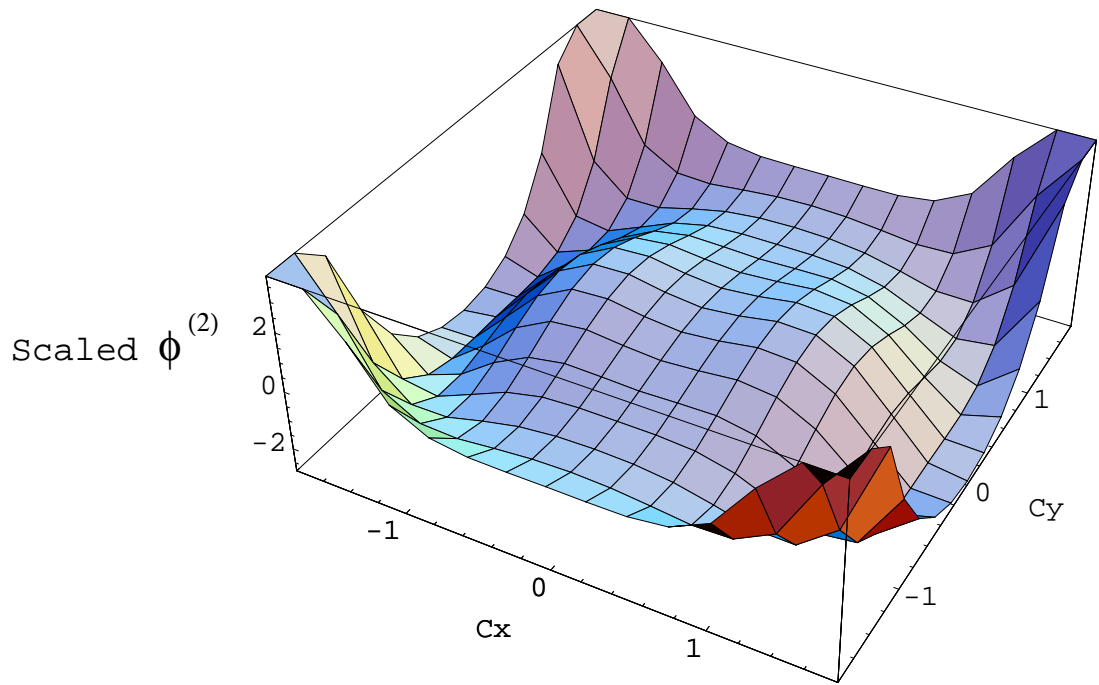} 
\caption{The scaled $\phi^{(2)}$ for hard-disk
 molecules with $7$th approximation $b_{0r}$ and
 $b_{2r}$.}
\label{f2cxcy}
\end{figure}

Figure \ref{MDf2} gives the $\phi^{(2)}$ in eq.(\ref{be46}) scaled by
$mJ_{x}^{2}/n^{2}\kappa^{3}T^{3}$ with the $3$th, $4$th, $5$th, $6$th and $7$th approximation $b_{0r}$ and $b_{2r}$. 
It should be mentioned that, as Fig.\ref{MDf2} shows, the scaled
$\phi^{(2)}$ in eq.(\ref{be46}) seems to converge to $7$th approximation. 
Figure \ref{f2cxcy} provides the explicit form of the scaled $\phi^{(2)}$ for hard-disk molecules with $7$th approximation $b_{0r}$ and $b_{2r}$. 
It is seen that the scaled $\phi^{(2)}$ for hard-disk molecules is strained symmetrically. 

\section{Other Nonequilibrium Velocity Distribution Functions to Second Order}\label{other}
\subsection{The Chapman-Enskog Solution of the two-dimensional Steady-State BGK Equation to 
 second order}\label{solvebgk}

For comparison, we also derive the velocity distribution
function for the two-dimensional steady-state BGK equation to second
 order by the Chapman-Enskog method.\cite{santos,santos1,santos4,santos2} 
Suppose a nonequilibrium system subject to a temperature gradient along the
$x$-axis in a steady
state whose velocity distribution
function is expressed as $f=f(x,{\bf v})$. 
The steady-state BGK equation is written as 
\begin{equation}  
v_{x} \partial_{x} f=\frac{f_{LE}-f}{\tau},\label{bgk10}
\end{equation}
where the relaxation time $\tau$ dependent on position $x$ through the density $n(x)$ and the
temperature $T(x)$.  
$f_{LE}$ is the usual local equilibrium 
velocity distribution function $f_{LE}(x,{\bf v})=(m n(x)/2\pi\kappa T(x))\exp[- m{\bf v}^{2}/2\kappa T(x)]$. 
It should be mentioned that, for the conservation laws, the collision term for the
steady-state BGK equation, the right-hand side of eq.(\ref{bgk10}), must satisfy 
\begin{equation}  
\int \Phi_{\mathrm{i}} f_{LE} d{\bf v}=\int \Phi_{\mathrm{i}} f d{\bf v},\label{bgk11}
\end{equation}
with the four collision invariants $\Phi_{\mathrm{i}}$ introduced in
eq.(\ref{be8}).
The velocity distribution function $f$ can be expanded as
\begin{equation}  
f=f^{(0)}+f^{(1)}+f^{(2)}+\cdots, \label{bgk30}
\end{equation} 
with $f^{(0)}\equiv f_{LE}$. 
Substituting eq.(\ref{bgk30}) into the steady-state BGK equation (\ref{bgk10}), we
arrive at the following set of equations:  
\begin{eqnarray}  
-\frac{f^{(1)}}{\tau}=v_{x} \partial_{x} f^{(0)}, \label{bgk40}
\end{eqnarray}
to first order and
\begin{eqnarray}  
-\frac{f^{(2)}}{\tau}=v_{x} \partial_{x} f^{(1)},\label{bgk50}
\end{eqnarray}
to second order. 
It is found that equation (\ref{bgk40}) with the requirement (\ref{bgk11}) leads to 
$n\kappa T$ being uniform, and that equation (\ref{bgk50}) with
the requirement (\ref{bgk11}) leads to heat flux $J_{x}$ calculated from eq.(\ref{be39A}) being uniform. 
Using these facts, from eqs.(\ref{bgk40}) and (\ref{bgk50}), the velocity distribution
function to second order for the the two-dimensional steady-state BGK equation becomes 
\begin{eqnarray} 
f&=&f^{(0)}\sum_{n=0}^{2}\left\{-\frac{J_{x}}{n\kappa T}\left(\frac{m}{2\kappa T}\right)^{\frac{1}{2}}\right\}^{n} n! c_{x}^{n} S_{1}^{n}({\bf c}^{2}),
\label{bgk110}
\end{eqnarray}
with the uniform heat flux
$J_{x}=-2n\kappa^{2}T\tau \partial_{x} T/m$. 

\subsection{Two-Dimensional Information Theory}\label{itsolve}

Let us construct two-dimensional information theory.\cite{jou,jou0}
The Zubarev form for the nonequilibrium velocity 
distribution function under a heat flux can be obtained by
maximizing the nonequilibrium entropy, defined as 
\begin{equation}  
S(x)\equiv-\kappa\int f\log f d{\bf v},\label{bgk130}
\end{equation}
under the constraints of the density:  
\begin{equation}  
n(x)\equiv\int f d{\bf v}
,\label{bgk51}
\end{equation} 
and the temperature:  
\begin{equation}  
n(x)\kappa T(x)\equiv
\int \frac{m{\bf v}^{2}}{2}f d{\bf v}.\label{bgk52}
\end{equation}
We assume no mean flow: 
\begin{equation} 
\int m {\bf v}f d{\bf v}={\bf 0}
,\label{bgk53}
\end{equation} 
where ${\bf 0}$ denotes the zero vector. 
Furthermore, we adopt the heat flux as a constraint:
\begin{equation}  
J_{x}\equiv\int \frac{m{\bf v}^{2}}{2}v_{x}f d{\bf v}. \label{bgk170} 
\end{equation}  
It should be emphasized that
 $n \kappa T$ and the heat flux $J_{x}$ are
now assumed to be uniform by contrast
with the case for the steady-state Boltzmann
equation where its solubility conditions lead to
$n \kappa T$ and $J_{x}$ being constant to second order. 

We have finally derived the nonequilibrium velocity 
distribution function to second order in the heat flux $J_{x}$ by
expanding the Zubarev's nonequilibrium velocity distribution
function to second order as
\begin{eqnarray}
f&=&
f^{(0)}\left\{1-
\frac{J_{x}}{n\kappa T}\left(\frac{m}{2\kappa T}\right)^{\frac{1}{2}}c_{x} S_{1}^{1}({\bf c}^{2})+\frac{mJ_{x}^{2}}{n^{2}\kappa^{3}T^{3}}\left(\frac{1}{2}-\frac{3{\bf c}^2}{4}\right)+
\frac{mJ_{x}^{2}}{n^{2}\kappa^{3}T^{3}}c_{x}^{2}
\left[1-{\bf c}^{2}+\frac{{\bf c}^{2}}{4}\right]\right\}. \nonumber \\
\label{bgk210} 
\end{eqnarray}  
In eq.(\ref{bgk210}), we have expanded the \textit{nonequilibrium temperature} which
has been obtained as
$\Theta=T(1-3mJ_{x}^{2}/4n^{2}\kappa^{3}T^{3})$ for
two dimension. 
Such the \textit{modified} velocity 
distribution function has been also obtained and used in three dimension.\cite{fort1,kim1} 
We note that, in all the macroscopic quantities calculated
in this paper, there are no differences between the results from the \textit{modified} velocity 
distribution function and Jou's velocity 
distribution function where \textit{nonequilibrium temperature} is not expanded\cite{jou,jou0}.  
Actually, the identifications of the density, the
temperature and the mean flow in eqs.(\ref{be10}), (\ref{be11}) and (\ref{be12}) do not affect the physical properties of the velocity
distribution function for the two-dimensional steady-state Boltzmann equation\cite{kim},  
and those identifications must be satisfied for the conservation laws 
in the case for the two-dimensional 
steady-state BGK equation. (see eq.(\ref{bgk11}))

\section{Direct comparison of the scaled $\phi^{(2)}$}\label{compare}

\begin{figure}[htbp]
\includegraphics[width=13.cm]{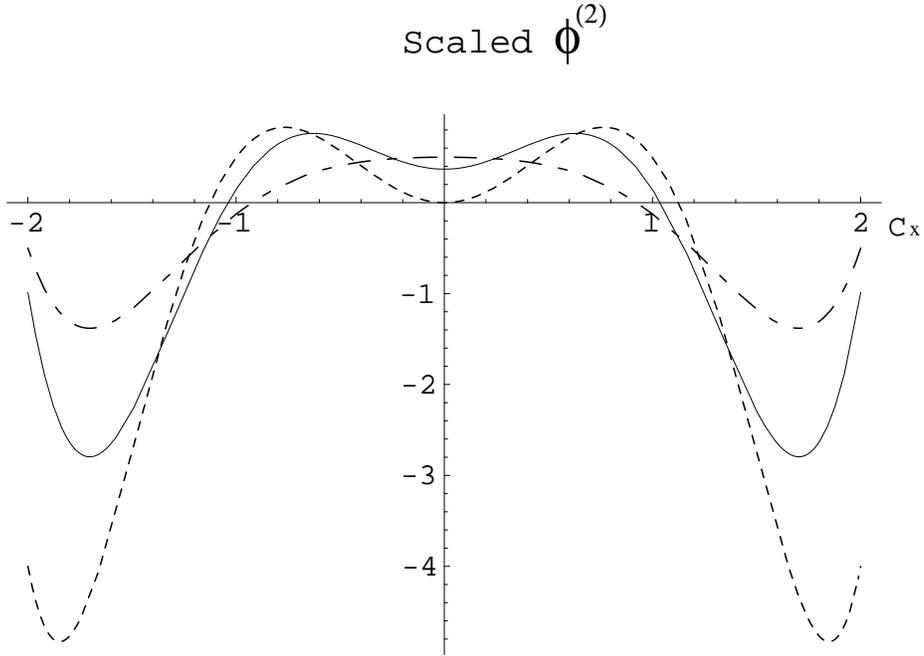} 
\caption{Direct comparison of the scaled $\phi^{(2)}$ for hard-disk
 molecules with those for the steady-state BGK
equation and information theory. 
The solid line, the dashed line and the dash-dotted line correspond to the scaled $\phi^{(2)}$ for hard-disk
 molecules with $7$th approximation $b_{0r}$ and
 $b_{2r}$, the steady-state BGK
equation and information theory, respectively. 
Note that we put $c_{y}=0$.}
\label{f2hikaku}
\end{figure}

Figure \ref{f2hikaku} exhibits the direct comparison of the scaled
$\phi^{(2)}$s for hard-disk molecules (\ref{be46}) to $7$th
approximation with those for the steady-state BGK
equation (\ref{bgk110}) and information theory (\ref{bgk210}). 
We have found that, as Fig.\ref{f2hikaku} explicitly shows, the second-order velocity distribution function for
 hard-disk molecules (\ref{be46}) definitely differs from the others. 
We emphasize that such a difference never appears to first order.

\section{Nonlinear Nonequilibrium Transport Phenomena}\label{application}
We can introduce the general form of the heat flux as 
\begin{eqnarray}  
J_{x}=-\varpi T^{\varphi} \partial_{x} T,    
\label{be470}
\end{eqnarray}
where $\varphi$ indicates temperature dependence of the thermal
conductivity and $\varpi$ is a constant that depends upon microscopic
models. 
For example, 
$\varphi$ is calculated as $1/2$ for hard-disk molecules, and $\varpi$ is determined as 
$2 b_{11} \kappa/d (\kappa/\pi m)^{1/2}$ with $b_{11}\simeq
 1.030$ for hard-disk molecules (see eq.(\ref{be39})). 
Note that $\varphi$ and $\varpi$ cannot be determined explicitly from the BGK
equation and information theory. 
From eq.(\ref{be470}), the temperature profile $T(x)$ in the
nonequilibrium steady state can be determined as
\begin{eqnarray}  
T(x)
&=&[T(0)^{\varphi+1}-(\varphi+1)\frac{J_{x}}{\varpi}x]^{\frac{1}{\varphi+1}}.
\label{be47}
\end{eqnarray}
It is seen that the temperature profile $T(x)$ becomes nonlinear except for
$\varphi=0$. 

Using eq.(\ref{be17.5}), the equation of state in the nonequilibrium steady
state can be obtained as
\begin{eqnarray}  
P_{ij}&=& n\kappa T[\delta_{ij}+\lambda_{P}^{ij}\frac{mJ_{x}^{2}}{n^{2}\kappa^{3}T^{3}}],\label{be51}
\end{eqnarray}
with the unit tensor $\delta_{ij}$ and the tensor components
$\lambda_{P}^{ij}$ given in Table \ref{macro1}. 
The values of $\lambda_{P}^{ij}$ for $7$th approximation $b_{1r}$,
$b_{0r}$ and $b_{2r}$ for hard-disk molecules seems to be converged to three significant figures, as can be seen from Table \ref{macro1}. 
Note that the off-diagonal components of $\lambda_{P}^{ij}$
are zero, and $\lambda_{P}^{xx}=-\lambda_{P}^{yy}$ is satisfied. 
Equation (\ref{be51}) shows that the equation of state in the nonequilibrium steady
state is not modified to first order. 
We indicate that the second-order pressure tensor $P_{ij}^{(2)}$ should be uniform from the solubility conditions for the third-order solution $\phi^{(3)}$:    
\begin{eqnarray}  
\int \Phi_{\mathrm{i}} {\bf v}_{\mathrm{1}}\cdot\nabla f^{(2)}_{\mathrm{1}} {\mathrm d}{\bf v}_{\mathrm{1}}=0.
\label{pconst}
\end{eqnarray}
Therefore, the pressure tensor $P_{ij}$ in eq.(\ref{be51}) becomes uniform since 
$n\kappa T$ is constant from the solubility conditions (\ref{be7}). 

We find that $\lambda_{P}^{ij}$ for hard-disk molecules differs from
that for the steady-state BGK equation and information theory not only quantitatively
  but also qualitatively:  $\lambda_{P}^{xx} < \lambda_{P}^{yy}$ for hard-disk molecules, while $\lambda_{P}^{xx}=\lambda_{P}^{yy}$ for the steady-state BGK equation and $\lambda_{P}^{xx} > \lambda_{P}^{yy}$ for information theory. 
This kind of the difference has been also found in the three-dimensional case.\cite{kim}

Each component of the kinetic temperature in the nonequilibrium steady
state, \textit{i.e.} $T_{\mathrm{i}}$ for ${\mathrm{i}}=x$ and $y$ is calculated as
\begin{eqnarray}  
\frac{n\kappa T_{\mathrm{i}}}{2}\equiv\left(\frac{2\kappa T}{m}\right)^{2}\int^{\infty}_{-\infty}{\mathrm d}{\bf c}\frac{mc_{\mathrm{i}}^{2}}{2}f,\end{eqnarray}
which leads to 
\begin{eqnarray}
T_{\mathrm{i}}=T[1+\lambda_{P}^{\mathrm{ii}}\frac{mJ_{x}^{2}}{n^{2}\kappa^{3}T^{3}}],\label{be54}
\end{eqnarray} 
for ${\mathrm{i}}=x$ and $y$. 
Values for the constants in the second-order term
$\lambda_{P}^{\mathrm{ii}}$ are the same as
$\lambda_{P}^{\mathrm{ij}}$ for $\mathrm{i}=\mathrm{j}$ given in Table \ref{macro1}. 
For hard-disk molecules, $T_{x}$ becomes smaller than $T_{y}$ regardless of the sign of $J_{x}$, which means that the
motion of hard-disk molecules along the heat flux becomes dull. 
We note that $T_{\mathrm{i}}$ for hard-disk molecules is isotropic to first order, that is, the equipartition law of energy holds.  

The Shannon entropy in the nonequilibrium steady
state $S$ is defined via
\begin{eqnarray}  
S&\equiv& -\frac{2\kappa^{2} T}{m}\int^{\infty}_{-\infty}{\mathrm d}{\bf c}f\log f \nonumber \\
&=&-n\kappa\log \left[\frac{n m}{2\pi \kappa T}\right]+n\kappa+\lambda_{S}\frac{mJ_{x}^{2}}{n\kappa^{2}T^{3}}.\label{be56}
\end{eqnarray}
Values for the constant $\lambda_{S}$ are given in Table \ref{macro1}: 
$\lambda_{S}$ for $7$th approximation $b_{1r}$, $b_{0r}$ and $b_{2r}$ for hard-disk
molecules seems to converge to four significant figures.  
It is found that $\lambda_{S}$ for hard-disk molecules is close to that for
the steady-state BGK equation and information theory. 
This is because the second-order correction term in the Shannon entropy is determined only by the square of the first-order solution $\phi^{(1)}$ where no important difference dependent on the kinetic equations or information theory appears. 
We note that the Shannon entropy in the nonequilibrium steady
state is not modified to first order.

\section{Contribution of the Steady Heat Conduction to the Rate of Chemical
Reaction}\label{contributionCR}
\subsection{Calculation of the Rate of Chemical Reaction}\label{CRsetudou}

In the early stage of a chemical reaction between monatomic gas molecules: 
\begin{equation}
A + A \rightarrow \mathrm{products}, \label{cr10}
\end{equation}
the rate of chemical reaction may not be affected by the existence of
products, and the reverse reaction can be neglected.\cite{pri} 
From the viewpoint of kinetic collision
theory, the chemical reaction rate (\ref{cr10}) can be described as 
\begin{eqnarray}  
R&=&
\int d{\bf v} \int d{\bf v}_{\mathrm{1}} \int d{\bf \Omega} \int f f_{\mathrm{1}} g \sigma(g),
\label{cr20}
\end{eqnarray}
for two dimension. 
Here ${\bf \Omega}$ denotes the solid angle for two dimension. 
For the differential cross-section of
chemical reaction $\sigma(g)$, we have derived the line-of-centers
model for the case of the two dimension.
Its form becomes
\begin{eqnarray}  
\sigma(g)= \left\{
\begin{array}{@{\,}ll}
0 & \quad g < \sqrt{\frac{4 E^{*}}{m}} \\
\frac{d}{\pi}\left(1-\frac{4 E^{*}}{m g^{2}}\right)^{\frac{1}{2}} & \quad g \ge \sqrt{\frac{4 E^{*}}{m}} \\
\end{array}
\right.,
\label{cr30}
\end{eqnarray} 
with $m$ mass of molecules and $E^{*}$ the
threshold energy of a chemical reaction. 

We calculate the rate of chemical reaction (\ref{cr20}) 
with the two-dimensional line-of-centers model (\ref{cr30})
using the explicit velocity distribution function of the two-dimensional steady-state Boltzmann equation for
 hard-core molecules to second order (\ref{be46}).
Substituting the expanded form of the velocity distribution function to
second order in eq.(\ref{be3}) into eq.(\ref{cr20}), we obtain
\begin{equation}  
R=R^{(0)}+R^{(1)}+R^{(2)}, \label{cr50}
\end{equation}
up to second order. 
The zeroth-order term of $R$,
\begin{equation}  
R^{(0)}=\int d{\bf v} \int d{\bf v}_{\mathrm{1}} \int d{\bf \Omega} \int f^{(0)} f_{\mathrm{1}}^{(0)} g \sigma(g)=2n^{2}d \left(\frac{\pi \kappa T}{m}\right)^{\frac{1}{2}}
e^{-\frac{E^{*}}{\kappa T}} , \label{cr60}
\end{equation}
corresponds to the rate of chemical reaction of the equilibrium theory. 
Similarly, the first-order term of $R$ is obtained as 
\begin{equation}  
R^{(1)}=\int d{\bf v} \int d{\bf v}_{\mathrm{1}} \int d{\bf \Omega} \int f^{(0)} f_{\mathrm{1}}^{(0)}[\phi^{(1)}+\phi^{(1)}_{\mathrm{1}}] g \sigma(g), \label{cr70}
\end{equation}
where $R^{(1)}$ does not appear because $\phi^{(1)}$ is an odd functions
of ${\bf c}$, as is shown in eq.(\ref{be46}).  
The second-order term of $R$, \textit{i.e.} $R^{(2)}$, is divided into 
\begin{equation}  
R^{(2,A)}=\int d{\bf v} \int d{\bf v}_{\mathrm{1}} \int d{\bf \Omega} \int f^{(0)} f_{\mathrm{1}}^{(0)}\phi^{(1)} \phi^{(1)}_{\mathrm{1}} g \sigma(g), \label{cr80}
\end{equation}
and
\begin{equation}  
R^{(2,B)}=\int d{\bf v} \int d{\bf v}_{\mathrm{1}} \int d{\bf \Omega} \int f^{(0)} f_{\mathrm{1}}^{(0)}[\phi^{(2)}+\phi^{(2)}_{\mathrm{1}}] g \sigma(g), \label{cr90}
\end{equation}
which exhibit the local nonequilibrium effect. 
Since the integrations (\ref{cr80}) and (\ref{cr90}) have the cutoff
form as in eq.(\ref{cr30}), the explicit forms of $\phi^{(1)}$ and $\phi^{(2)}$ of the steady-state Boltzmann equation for
hard-disk molecules are required to calculate $R^{(2,A)}$ and
$R^{(2,B)}$, respectively. 

In Fig. \ref{MDf2}, we have confirmed that $\phi^{(2)}$ seems to converge to $7$th Sonine
approximation, so that we will show only the
results calculated from $\phi^{(1)}$ and $\phi^{(2)}$ for $7$th approximation of Sonine polynomials. 
In order to compare the results from the steady-state Boltzmann equation with those from the
 steady-state BGK equation and information theory, we also use the 
 explicit forms of $\phi^{(1)}$ and $\phi^{(2)}$ obtained in eq.(\ref{bgk110})
 for the steady-state
BGK equation and in eq.(\ref{bgk210}) for
information theory. 

\subsection{Local Nonequilibrium Effect on the Rate of Chemical Reaction}\label{CRresults}
Inserting $\phi^{(1)}$ and $\phi^{(2)}$ of
eq.(\ref{be46}) for the steady-state Boltzmann equation for hard-core molecules,
eq.(\ref{bgk110}) for the steady-state BGK equation and
eq.(\ref{bgk210}) for
information theory into eqs.(\ref{cr80}) and (\ref{cr90}), and
performing the integrations with the chemical reaction cross-section
(\ref{cr30}), we finally obtain the local nonequilibrium effect on
the rate of chemical reaction based on the line-of-centers model. 
The expressions of $R^{(2,A)}$ and $R^{(2,B)}$ become
\begin{eqnarray}  
R^{(2,A)}=\frac{d m J_{x}^{2}}{\kappa^{3}T^{3}}\left(\frac{\pi \kappa T}{m}\right)^{\frac{1}{2}}
e^{-\frac{E^{*}}{\kappa T}} \{\sum_{r\ge 0} \alpha_{r} \left(\frac{E^{*}}{\kappa T}\right)^{r}\}
, 
\label{cr120}
\end{eqnarray}
and
\begin{eqnarray}  
R^{(2,B)}=\frac{d m J_{x}^{2}}{\kappa^{3}T^{3}}\left(\frac{\pi \kappa T}{m}\right)^{\frac{1}{2}}
e^{-\frac{E^{*}}{\kappa T}} \{\sum_{r\ge 0} \beta_{r} \left(\frac{E^{*}}{\kappa T}\right)^{r}\}
, 
\label{cr130}
\end{eqnarray}
respectively. 
The numerical values for $\alpha_{r}$ and $\beta_{r}$ are listed in
Tables \ref{alpha} and \ref{beta}, respectively. 
As well as the three dimension case, the two-dimensional $R^{(2,B)}$ for steady-state Boltzmann equation is determined only by the terms of
$b_{0r}$ in $\phi^{(2)}$ of eq.(\ref{be46}). 
This is because $R^{(2,B)}$ of eq.(\ref{cr90}) has $x$-$y$ symmetry, so that the terms including $(c_{x}^2-c_{y}^2)$ in $\phi^{(2)}$ of eq.(\ref{be46}) do not contribute to $R^{(2,B)}$.

\begin{figure}[htbp]
\includegraphics[width=13.cm]{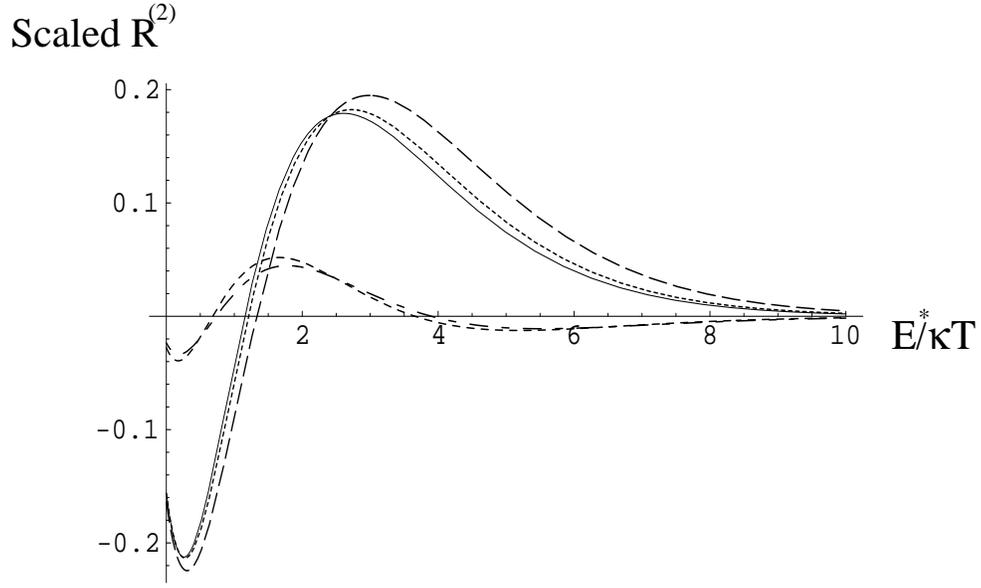} 
\caption{Scaled $R^{(2)}$ compared to scaled $R^{(2,A)}$ as a function
 of $E^{*}/\kappa T$ for the two-dimensional case.
The solid line, the long-dashed line and the dotted line show $R^{(2)}$ for hard-disk molecules, the steady-state BGK equation and information theory, respectively.The dashed line and the dash-dotted line represent $R^{(2,A)}$ for hard-disk molecules and both of the steady-state BGK equation and information theory, respectively.}
\label{CR}
\end{figure}

The graphical results of $R^{(2)}$ compared with those of $R^{(2,A)}$
are provided in Fig.\ref{CR}. 
Both of $R^{(2)}$ and $R^{(2,A)}$ in Fig.\ref{CR} are scaled by $\pi^{1/2}d m^{1/2}
J_{x}^{2}/\kappa^{5/2}T^{5/2}$. 
Note that $R^{(2)}$ is the sum of $R^{(2,A)}$ and $R^{(2,B)}$ in eqs.(\ref{cr120}) and
(\ref{cr130}). 
As Fig.\ref{CR} shows, it is clear that $R^{(2,B)}$ plays an essential role for the evaluation
of $R^{(2)}$.
We have found that there are no qualitative differences among $R^{(2)}$ and $R^{(2,A)}$ for the steady-state Boltzmann equation, and those for the steady-state BGK equation and information theory, while they exhibit slight deviations from each other. 
The quantitative deviation in $R^{(2,A)}$ would not be observed if we adopted
$\phi^{(1)}$ of the steady-state Boltzmann equation for 
the lowest Sonine approximation, because that is identical with the precise $\phi^{(1)}$ of the steady-state BGK
 equation and information theory. 

\section{Discussion and Conclusion}\label{ktdiscussion}

Fushiki has recently demonstrated that our analytical three-dimensional second-order solution of the
steady-state Boltzmann equation for hard-core molecules agrees well with results of his numerical
experiment using both a molecular dynamics simulation and a direct simulation monte carlo method.\cite{fushiki,kimdoctor} 
Using the method developed in ref.\cite{kim}, we have derived the velocity distribution function of the two-dimensional steady-state Boltzmann equation for
 hard-disk molecules explicitly to second order in the temperature gradient, as was shown explicitly in eq.(\ref{be46}) and graphically in Fig.\ref{f2cxcy}. 
We have calculated the two-dimensional equation of state for hard-disk molecules to second order from it. 
We believe that the second-order solution of the steady-state Boltzmann
equation is physically important in that it reflects a nonequilibrium state far from equilibrium, called the local nonequilibrium state. 

We have found that there are qualitative differences between hard-disk molecules and 
the steady-state BGK equation
in the nonlinear nonequilibrium transport phenomena based on the local nonequilibrium state:
the second-order corrections appear for hard-disk molecules in the pressure tensor $P_{ij}$
and the kinetic temperature $T_{\mathrm{i}}$, while no corrections to these
 quantities appear for the steady-state BGK equation, as
 Table \ref{macro1} shows. 
This kind of qualitative differences was detected also in the three-dimensional case.\cite{kim}
This discrepancy is due to 
the fact that $g$-dependency cannot be absorbed in the
single relaxation time of the BGK equation, which leads to the conclusion that the steady-state BGK
equation neither capture the essence of hard-disk molecules nor possess the characteristics of any other models of molecules which interact with $g$-dependency. 
We suggest that microscopic models which possess the property
that its relaxation to the local equilibrium state is
 described only by a single relaxation time 
 could not be applied to describe the nonlinear nonequilibrium transport phenomena. 
This suggestion may mean that the steady-state BGK equation could capture the essence
 of hard-disk molecules if one made the relaxation time depend 
 on $g$ or if one developed the steady-state BGK equation with
 multi-relaxation times. 
We note that the qualitative differences mentioned above still appear no matter which boundary condition is adopted, that is, the isotropy and  
the anisotropy of the pressure tensor in eq.(\ref{be51})
and the kinetic temperature in eq.(\ref{be54}) are not affected by any
 kinds of boundary conditions. 

We have examined information theory by the microscopic kinetic theory mentioned above, and consider the possibility of the existence of a nonequilibrium universal velocity distribution
function. 
The first-order velocity distribution
 functions for the steady-state
Boltzmann equation for hard-disk molecules, \textit{i.e.} the first-order terms in eq.(\ref{be46}), is
consistent with that derived by expanding Zubarev's velocity distribution
function\cite{zubarev,zubarev1,katz}.  
On the other hand, the explicit form of the second-order term in eq.(\ref{be46}) definitely differs from the precise
 form for the steady-state BGK equation (\ref{bgk110}) or information theory (\ref{bgk210}),
as Fig.\ref{f2hikaku} shows. 
Although information theory has been applied 
 to nonequilibrium dilute
gases\cite{fort,fort1,jou1,jou4,jou2,jou3}, we have found that information theory contradicts the microscopic kinetic models: all the macroscopic quantities for information
theory except for the
Shannon entropy $S$ in eq.(\ref{be56}) are qualitatively different from those for the steady-state
Boltzmann equation and the steady-state BGK equation. 
These results indicate that characteristics of microscopic models appear in the local nonequilibrium state, that is, nonlinear nonequilibrium transport phenomena are sensitive to differences of kinetic models, so rather realistic models are needed when one investigates them.  
We can conclude that, though quite a few statistical physicists have believed the existence of a universal velocity distribution
function in the nonequilibrium steady state by maximizing the
Shannon-type entropy\cite{zubarev,onsager,eu,nettleton1,jou,zubarev1,katz}, any universal nonlinear nonequilibrium velocity distribution
function does not seem to exist in the two-dimensional case as well as the three-dimensional case, even when it is expressed only in terms of macroscopic quantities. 
We suggest that the entropy defined in eq.(\ref{bgk130})
is not appropriate as the nonequilibrium entropy to second order though it is appropriate to first order, and
that some
nonequilibrium corrections dependent on microscopic models are needed for 
the nonequilibrium entropy to second order.   

The second-order solution of the steady-state Boltzmann equation for
 hard-disk molecules is indispensable for the calculation of the nonequilibrium
 effects on the rate of chemical reaction, since $R^{(1)}$
 does not appear and $R^{(2,B)}$ is remarkably
 larger than $R^{(2,A)}$ as Fig.\ref{CR} shows. 
This indicates the significance of the second-order coefficients 
 as terms which reflect the local nonequilibrium state.  

\section{Acknowledgments}

I thank H. Hayakawa for useful discussions. 
This research was partially supported by the Japan Science Society.

\appendix
\section{Calculation of $\Omega_{\lowercase{kr}}$}\label{a3}
From the definition of $Q_{kr}$, $\Omega_{kr}$ can be
calculated using the mathematical properties of the trigonometrical functions and Sonine polynomials.
For example, $c_{{\mathrm{1}} x} Q_{kr}$ can be rewritten as 
\begin{equation}  
\left(\frac{2\kappa T}{m}\right)^{\frac{1}{2}}c_{{\mathrm{1}} x} Q_{kr}=2(\delta_{k,0}+1)\left(\frac{m}{2\kappa T}\right)^{\frac{k}{2}}\frac{S^{r}_{k}({\bf c}^{2}_{\mathrm{1}})}{\Gamma(k+r+1)}[Y_{k+1}({\bf c}_{\mathrm{1}})+\frac{2\kappa T}{m}c^{2}_{\mathrm{1}}Y_{k-1}({\bf c}_{\mathrm{1}})].\label{be21}
\end{equation}
Integrating eq.(\ref{be21}) over $(2\kappa T/m)^{1/2}{\bf c}_{\mathrm{1}}$ with $f_{\mathrm{1}}$
from eq.(\ref{be13}) can be performed by using the following orthogonality properties. 
For Sonine
polynomials, 
\begin{eqnarray}  
\int_{0}^{\infty} X^{k} e^{-X} S_{k-1}^{p}(X)S_{k}^{q}(X) dX=\frac{(-1)^{p-q}\Gamma(q+k+1)}{q!},
\label{be13.8}
\end{eqnarray}
for $p=q$ and $p=q+1$, and is zero otherwise. 
For the trigonometrical functions, 
\begin{eqnarray}  
\int_{0}^{2\pi} \cos n\phi \sin m\phi d\phi=0,
\label{be18.5}
\end{eqnarray}
and
\begin{eqnarray}  
\int_{0}^{2\pi} \cos n\phi \cos m\phi d\phi=\int_{0}^{2\pi} \sin n\phi \sin m\phi d\phi=\pi \delta_{nm},
\label{be18.51}
\end{eqnarray}
with the Kronecker delta $\delta_{pq}$ for $m \ne 0$. 
We can calculate $D_{k,r}$ and
$E_{k,r}$ defined as   
\begin{eqnarray}  
D_{k,r}\equiv \frac{1}{n}\overline{c_{{\mathrm{1}} x} Q_{kr}} \quad {\mathrm and} \quad E_{k,r}\equiv\frac{1}{n}\overline{c_{{\mathrm{1}} y} Q_{kr}}. 
\end{eqnarray}
The results can be written as 
\begin{eqnarray}
D_{k,r}&=&(\delta_{k,0}+1)\left[(k+r+1)B_{k+1,r}-B_{k+1,r-1}+(\delta_{k,0}+1)B_{k-1,r}-(\delta_{k,0}+1)(r+1)B_{k-1,r+1}\right], \nonumber \\ 
\label{be22}
\end{eqnarray}
and 
\begin{eqnarray}  
E_{k,r}&=&(\delta_{k,0}+1)\left[(k+r+1)C_{k+1,r}-C_{k+1,r-1}-C_{k-1,r}+(r+1)C_{k-1,r+1}\right]. \nonumber \\ 
\label{be23}
\end{eqnarray}
Additionally, $c_{{\mathrm{1}} x} \partial_{x} Q_{kr}$ can be rewritten as 
\begin{equation}  
c_{{\mathrm{1}} x} \partial_{x} Q_{kr}=\frac{\partial_{x} T}{T}c_{{\mathrm{1}} x}\left[Q_{k,r-1}({\bf c}_{\mathrm{1}})-(r+\frac{k}{2})Q_{k,r}\right],\label{be21.5}
\end{equation}
Therefore, by integrating eq.(\ref{be21.5}) over $(2\kappa T/m)^{1/2}{\bf c}_{\mathrm{1}}$, with $f_{\mathrm{1}}$
from eq.(\ref{be13}), it is found that 
\begin{eqnarray}  
\overline{c_{{\mathrm{1}} x} \partial_{x} Q_{kr}}=n\frac{\partial_{x} T}{T}\left[D_{k,r-1}-(r+\frac{k}{2})D_{k,r}\right].\label{be25}
\end{eqnarray}
Similarly 
$\overline{c_{{\mathrm{1}} y} \partial_{y} Q_{kr}}$ is obtained by replacing the
differential coefficients with respect to $x$ by the corresponding 
differential coefficients with respect to $y$, the $D_{k,r}$'s by the
corresponding $E_{k,r}$, respectively.  
Substituting these results into eq.(\ref{be19.0}), $\Omega_{kr}$ finally
becomes eq.(\ref{be25.5}). 

\section{Calculation of $F_{kr}^{1}(\chi)$}
\subsection{Details of $F_{kr}^{1}(\chi)$}\label{a2}
The details of $F_{kr}^{1}(\chi)$ are written in this Appendix.  
Substituting the general forms of $f_{\mathrm{1}}$, $f_{\mathrm{2}}$ in eq.(\ref{be13}) 
and $Q^{\prime}_{kr}$ in eq.(\ref{be20}) into $F_{kr}^{1}(\chi)$ in
eq.(\ref{be27}), $F_{kr}^{1}(\chi)$ can be written as
\begin{eqnarray}
F_{kr}^{1}(\chi)=\sum_{n_{\mathrm{1}}, n_{\mathrm{2}}, k_{\mathrm{1}}, k_{\mathrm{2}}}W_{k,k_{\mathrm{1}},k_{\mathrm{2}}}^{r,n_{\mathrm{1}}, n_{\mathrm{2}}} \left\{\Xi_{k,k_{\mathrm{1}},k_{\mathrm{2}}}^{Y,r,n_{\mathrm{1}},n_{\mathrm{2}}}(\chi) B_{k_{\mathrm{1}}n_{\mathrm{1}}}B_{k_{\mathrm{2}}n_{\mathrm{2}}}+\Xi_{k,k_{\mathrm{1}},k_{\mathrm{2}}}^{Z,r,n_{\mathrm{1}},n_{\mathrm{2}}}(\chi) C_{k_{\mathrm{1}}n_{\mathrm{1}}}C_{k_{\mathrm{2}}n_{\mathrm{2}}}\right\},\label{be34.5}
\end{eqnarray}
where $\Xi_{k,k_{\mathrm{1}},k_{\mathrm{2}}}^{Y,r,n_{\mathrm{1}},n_{\mathrm{2}}}(\chi)$
is the characteristic integral defined as 
\begin{eqnarray}
\Xi_{k,k_{\mathrm{1}},k_{\mathrm{2}}}^{Y,r,n_{\mathrm{1}},n_{\mathrm{2}}}(\chi)&\equiv& \int \int \exp[-(c_{\mathrm{1}}^{2}+c^{2}_{\mathrm{2}})] Y_{k}({\bf c}_{\mathrm{1}}^{\prime}) Y_{k_{\mathrm{1}}}({\bf c}_{\mathrm{1}})Y_{k_{\mathrm{2}}}({\bf c}_{\mathrm{2}}) S^{r}_{k}({\bf c}_{\mathrm{1}}^{\prime 2})S^{n_{\mathrm{1}}}_{k_{\mathrm{1}}}({\bf c}_{\mathrm{1}}^{2})S^{n_{\mathrm{2}}}_{k_{\mathrm{2}}}({\bf c}_{\mathrm{2}}^{2}) g {\mathrm d}{\bf c}_{\mathrm{2}}{\mathrm d}{\bf c}_{\mathrm{1}},\nonumber\\
\label{be29}
\end{eqnarray}
and
\begin{eqnarray}
\Xi_{k,k_{\mathrm{1}},k_{\mathrm{2}}}^{Z,r,n_{\mathrm{1}},n_{\mathrm{2}}}(\chi)&\equiv& \int \int \exp[-(c_{\mathrm{1}}^{2}+c^{2}_{\mathrm{2}})] Y_{k}({\bf c}_{\mathrm{1}}^{\prime}) Z_{k_{\mathrm{1}}}({\bf c}_{\mathrm{1}})Z_{k_{\mathrm{2}}}({\bf c}_{\mathrm{2}}) S^{r}_{k}({\bf c}_{\mathrm{1}}^{\prime 2})S^{n_{\mathrm{1}}}_{k_{\mathrm{1}}}({\bf c}_{\mathrm{1}}^{2})S^{n_{\mathrm{2}}}_{k_{\mathrm{2}}}({\bf c}_{\mathrm{2}}^{2}) g {\mathrm d}{\bf c}_{\mathrm{2}}{\mathrm d}{\bf c}_{\mathrm{1}}.\nonumber\\
\label{be29Z}
\end{eqnarray}
We note that values of the latter is obtained from those of the former by a transformation of axes, and that $\Xi_{k,0,k_{\mathrm{2}}}^{Z,r,n_{\mathrm{1}},n_{\mathrm{2}}}(\chi)=\Xi_{k,k_{\mathrm{1}},0}^{Z,r,n_{\mathrm{1}},n_{\mathrm{2}}}(\chi)=0$ from
eq.(\ref{be17}).  
The integral containing $Y_{k_{\mathrm{1}}}({\bf
c}_{\mathrm{1}})Z_{k_{\mathrm{2}}}({\bf c}_{\mathrm{2}})+Z_{k_{\mathrm{1}}}({\bf
c}_{\mathrm{1}})Y_{k_{\mathrm{2}}}({\bf c}_{\mathrm{2}})$ becomes
zero, owing to the symmetry of the trigonometrical functions. 
The factor $W_{k,k_{\mathrm{1}},k_{\mathrm{2}}}^{r,n_{\mathrm{1}},
n_{\mathrm{2}}}$ in eq.(\ref{be34.5}) is defined as
\begin{eqnarray}
W_{k,k_{\mathrm{1}},k_{\mathrm{2}}}^{n_{\mathrm{1}}, n_{\mathrm{2}}}\equiv \frac{4 n^{2}}{\pi^{2}\Gamma(k+r+1)}\left(\frac{m}{2\kappa T}\right)^\frac{k+k_{\mathrm{1}}+k_{\mathrm{2}}}{2} n_{\mathrm{1}}! n_{\mathrm{2}}! , \label{be34}
\end{eqnarray}
which is obtained from the prefactors and the coefficients in the general form of
$f_{\mathrm{1}}$, $f_{\mathrm{2}}$ in eq.(\ref{be13}) and
$Q^{\prime}_{kr}$ in eq.(\ref{be20}). 

We find that it is necessary only to evaluate the characteristic integral
$\Xi_{k,k_{\mathrm{1}},k_{\mathrm{2}}}^{Y,r,n_{\mathrm{1}},n_{\mathrm{2}}}(\chi)$
in order to calculate $F_{kr}^{1}(\chi)$. 
Our calculation of
$\Xi_{k,k_{\mathrm{1}},k_{\mathrm{2}}}^{Y,r,n_{\mathrm{1}},n_{\mathrm{2}}}(\chi)$
is written in the next subsection. 
Once the characteristic integral $\Xi_{k,
k_{\mathrm{1}},k_{\mathrm{2}}}^{Y,r,n_{\mathrm{1}},
n_{\mathrm{2}}}(\chi)$ has been derived,
$F_{kr}^{1}(\chi)$ is now calculated from eq.(\ref{be34.5}) with
$W_{k,k_{\mathrm{1}},k_{\mathrm{2}}}^{r,n_{\mathrm{1}},
n_{\mathrm{2}}}$ in eq.(\ref{be34}). 

\subsection{Calculation of $\Xi_{k,k_{\mathrm{1}},k_{\mathrm{2}}}^{Y,r,n_{\mathrm{1}},n_{\mathrm{2}}}(\lowercase{\chi})$}\label{kimref}

We shall explain how to calculate
 $\Xi_{\lowercase{k,k_{\mathrm{1}},k_{\mathrm{2}}}}^{Y,r,n_{\mathrm{1}},n_{\mathrm{2}}}(\lowercase{\chi})$ which appears in eq.(\ref{be29}). 
The calculation has been performed mainly based on the method developed in
ref.\cite{kim}.  

\subsubsection{Introduction of $\Theta_{k,k_{\mathrm{1}},k_{\mathrm{2}}}$}
Using eq.(\ref{be13.5}), the characteristic integral $\Xi_{k,k_{\mathrm{1}},k_{\mathrm{2}}}^{Y,r,n_{\mathrm{1}},n_{\mathrm{2}}}(\chi)$ corresponds to the coefficient of $s^{r}t^{n_{\mathrm{1}}}u^{n_{\mathrm{2}}}$ in
\begin{eqnarray}
\Theta_{k,k_{\mathrm{1}},k_{\mathrm{2}}}&\equiv&\nu_{k,k_{\mathrm{1}},k_{\mathrm{2}}}\int \int \tilde{Y}_{k}({\bf c}_{\mathrm{1}}^{\prime}) \tilde{Y}_{k_{\mathrm{1}}}({\bf c}_{\mathrm{1}})\tilde{Y}_{k_{\mathrm{2}}}({\bf c}_{\mathrm{2}})\exp\{-(\alpha c^{2}_{\mathrm{1}}+ \beta c^{2}_{\mathrm{2}}+\gamma c^{\prime 2}_{\mathrm{1}})\} g d{\bf c}_{\mathrm{2}}d{\bf c}_{\mathrm{1}},
\label{be31}
\end{eqnarray}
that is,  
\begin{eqnarray}
\Theta_{k,k_{\mathrm{1}},k_{\mathrm{2}}}\equiv\sum_{r,n_{\mathrm{1}},n_{\mathrm{2}}}\Xi_{k,k_{\mathrm{1}},k_{\mathrm{2}}}^{Y,r,n_{\mathrm{1}},n_{\mathrm{2}}}(\chi) s^{r}t^{n_{\mathrm{1}}}u^{n_{\mathrm{2}}}. 
\label{be31.5}
\end{eqnarray}
In eq.(\ref{be31}) $\alpha$, $\beta$ and $\gamma$ are defined as
\begin{eqnarray}
\alpha\equiv \frac{1}{1-t}, \quad \beta\equiv \frac{1}{1-u}, \quad \gamma\equiv\frac{s}{1-s}, \label{be32.1}
\end{eqnarray}
and $\nu_{k,k_{\mathrm{1}},k_{\mathrm{2}}}$ is given by 
\begin{eqnarray}
\nu_{k,k_{\mathrm{1}},k_{\mathrm{2}}}\equiv (1-s)^{-k-1}(1-t)^{-k_{\mathrm{1}}-1}(1-u)^{-k_{\mathrm{2}}-1} \left(\frac{2\kappa T}{m}\right)^{\frac{k+k_{\mathrm{1}}+k_{\mathrm{2}}}{2}},  \label{be32.5}
\end{eqnarray}
and 
\begin{eqnarray}
\tilde{Y}_{k}({\bf c}_{\mathrm{1}}^{\prime})=c_{\mathrm{1}}^{\prime k} \cos k \phi ,\quad  \tilde{Y}_{k_{\mathrm{1}}}({\bf c}_{\mathrm{1}})=c_{\mathrm{1}}^{k_{\mathrm{1}}} \cos k_{\mathrm{1}} \phi_{\mathrm{1}} \quad \mathrm{and} \quad  \tilde{Y}_{k_{\mathrm{2}}}({\bf c}_{\mathrm{2}})=c_{\mathrm{2}}^{k_{\mathrm{2}}} \cos k_{\mathrm{2}} \phi_{\mathrm{2}}. 
\label{be31-0}
\end{eqnarray}
Finally, we need only to evaluate the characteristic integral
$\Theta_{k,k_{\mathrm{1}},k_{\mathrm{2}}}$
in order to calculate $\Xi_{k,k_{\mathrm{1}},k_{\mathrm{2}}}^{Y,r,n_{\mathrm{1}},n_{\mathrm{2}}}(\chi)$. 

\subsubsection{Derivation of the Inductive Equation}
In order to evaluate
$\Theta_{k,k_{\mathrm{1}},k_{\mathrm{2}}}$
in eq.(\ref{be31}), let us
derive an inductive equation for 
$\tilde{\Theta}_{k,k_{\mathrm{1}},k_{\mathrm{2}}}$, which is related to
$\Theta_{k,k_{\mathrm{1}},k_{\mathrm{2}}}$
by 
\begin{eqnarray}
\tilde{\Theta}_{k,k_{\mathrm{1}},k_{\mathrm{2}}}
\equiv\nu_{k,k_{\mathrm{1}},k_{\mathrm{2}}}^{-1}\Theta_{k,k_{\mathrm{1}},k_{\mathrm{2}}}.    
\label{be401}
\end{eqnarray}
By replacing $c_{{\mathrm{1}}x}$ and $c_{{\mathrm{2}}x}$ by
$c_{{\mathrm{1}}x}-w$ and
$c_{{\mathrm{2}}x}-w$, respectively, $c^{\prime}_{{\mathrm{1}}x}$ and
$c^{\prime}_{{\mathrm{2}}x}$ will be changed to $c^{\prime}_{{\mathrm{1}}x}-w$
and $c^{\prime}_{{\mathrm{2}}x}-w$. 
At the same time, the relative speed $g$ is not modified, and the
value of $\Theta_{k,k_{\mathrm{1}},k_{\mathrm{2}}}$ is unchanged. 
Therefore, $\Theta_{k,k_{\mathrm{1}},k_{\mathrm{2}}}$ is independent of $w$ and differentiation of
$\Theta_{k,k_{\mathrm{1}},k_{\mathrm{2}}}$ with respect to $w$ gives zero. 
After this differentiation has been performed and $w$ has been set to zero, it is found that
\begin{eqnarray}
& &\int \int \exp\{-(\alpha c^{2}_{\mathrm{1}}+ \beta c^{2}_{\mathrm{2}}+\gamma c^{\prime 2}_{\mathrm{1}})\} g d{\bf c}_{\mathrm{2}}d{\bf c}_{\mathrm{1}}\times \nonumber \\
& &\left\{ \alpha \tilde{Y}_{k} \tilde{Y}_{k_{\mathrm{2}}} (\tilde{Y}_{k_{\mathrm{1}}+1}+ c^{2}_{\mathrm{1}} \tilde{Y}_{k_{\mathrm{1}}-1}) (\delta_{k_{\mathrm{1}},0}+1)
+\beta \tilde{Y}_{k} \tilde{Y}_{k_{\mathrm{1}}} (\tilde{Y}_{k_{\mathrm{2}}+1}+ c^{2}_{\mathrm{2}} \tilde{Y}_{k_{\mathrm{2}}-1}) (\delta_{k_{\mathrm{2}},0}+1)\right.\nonumber \\
& &\left. +\gamma \tilde{Y}_{k_{\mathrm{1}}} \tilde{Y}_{k_{\mathrm{2}}} (\tilde{Y}_{k+1}+ c^{\prime 2}_{\mathrm{1}} \tilde{Y}_{k-1}) (\delta_{k,0}+1) -k_{\mathrm{1}} \tilde{Y}_{k}\tilde{Y}_{k_{\mathrm{1}}-1}\tilde{Y}_{k_{\mathrm{2}}}-k_{\mathrm{2}}\tilde{Y}_{k}\tilde{Y}_{k_{\mathrm{1}}} \tilde{Y}_{k_{\mathrm{2}}-1}-k \tilde{Y}_{k-1}\tilde{Y}_{k_{\mathrm{1}}}\tilde{Y}_{k_{\mathrm{2}}}\right\}=0, \nonumber \\
\label{be32}
\end{eqnarray}
by using the formulae
\begin{equation}  
 c_{{\mathrm{i}} x} \tilde{Y}_{k_{\mathrm{i}}}({\bf c}_{\mathrm{i}}) =\frac{(\delta_{k_{{\mathrm{i}}},0}+1)}{2} \left\{\tilde{Y}_{k_{{\mathrm{i}}}+1}({\bf c}_{\mathrm{i}})+ c^{2}_{\mathrm{i}}\tilde{Y}_{k_{{\mathrm{i}}}-1}({\bf c}_{\mathrm{i}})\right\},
\label{be214}
\end{equation}
and 
\begin{equation}  
\frac{\partial \tilde{Y}_{k_{{\mathrm{i}}}}({\bf c}_{{\mathrm{i}}})}{\partial c_{{\mathrm{i}}x}}=k_{{\mathrm{i}}} \tilde{Y}_{k_{{\mathrm{i}}}-1}({\bf c}_{{\mathrm{i}}}), 
\label{be215}
\end{equation}
for $\mathrm{i}=1$ and $2$.
From eqs.(\ref{be31}) and (\ref{be401}), eq.(\ref{be32}) leads to the inductive equation 
\begin{eqnarray}
& &\alpha (\delta_{k_{\mathrm{1}},0}+1) \tilde{\Theta}_{k,k_{\mathrm{1}}+1,k_{\mathrm{2}}} 
-\alpha (\delta_{k_{\mathrm{1}},0}+1) \frac{\partial \tilde{\Theta}_{k,k_{\mathrm{1}}-1,k_{\mathrm{2}}}}{\partial \alpha} \nonumber \\
&+&\beta (\delta_{k_{\mathrm{2}},0}+1) \tilde{\Theta}_{k,k_{\mathrm{1}},k_{\mathrm{2}}+1}
-\beta (\delta_{k_{\mathrm{2}},0}+1) \frac{\partial \tilde{\Theta}_{k,k_{\mathrm{1}},k_{\mathrm{2}}-1}}{\partial \beta} \nonumber \\
&+&\gamma (\delta_{k,0}+1) \tilde{\Theta}_{k+1,k_{\mathrm{1}},k_{\mathrm{2}}}
-\gamma (\delta_{k,0}+1) \frac{\partial \tilde{\Theta}_{k-1,k_{\mathrm{1}},k_{\mathrm{2}}}}{\partial \gamma} \nonumber \\
&-&k\tilde{\Theta}_{k-1,k_{\mathrm{1}},k_{\mathrm{2}}}-k_{\mathrm{1}}\tilde{\Theta}_{k,k_{\mathrm{1}}-1,k_{\mathrm{2}}}-k_{\mathrm{2}}\tilde{\Theta}_{k,k_{\mathrm{1}},k_{\mathrm{2}}-1}
=0,  \nonumber \\
\label{be320}
\end{eqnarray}
for
$\tilde{\Theta}_{k,k_{\mathrm{1}},k_{\mathrm{2}}}$. 
Because of this inductive equation, once the initial value $\tilde{\Theta}_{0,k_{\mathrm{1}},k_{\mathrm{2}}}$
is known for all $k_{\mathrm{1}}$ and $k_{\mathrm{2}}$, then the values of the integral
$\tilde{\Theta}_{k,k_{\mathrm{1}},k_{\mathrm{2}}}$
for any $k$, $k_{\mathrm{1}}$ and $k_{\mathrm{2}}$ can be obtained, and     
$\Theta_{k,k_{\mathrm{1}},k_{\mathrm{2}}}$
is then obtained from eq.(\ref{be401}). 

\subsubsection{Calculation of the Initial Value}\label{init}

In principle, the initial value of the inductive equation (\ref{be320}),
$\tilde{\Theta}_{0,k_{\mathrm{1}},k_{\mathrm{2}}}$, 
can be obtained and written explicitly from eq.(\ref{be31}), changing the variables ${\bf c}_{{\mathrm{1}}}$ and ${\bf c}_{{\mathrm{2}}}$ to ${\bf V}=({\bf c}_{{\mathrm{1}}}+{\bf c}_{{\mathrm{2}}})/2$ and ${\bf g}={\bf c}_{{\mathrm{1}}}-{\bf c}_{{\mathrm{2}}}$. 
Though we have directly calculated $\tilde{\Theta}_{0,k_{\mathrm{1}},k_{\mathrm{2}}}$ only for $k_{\mathrm{1}}=k_{\mathrm{2}}=0$, $1$, $2$, $3$, $4$, $5$, $6$, $7$ and $8$, they are sufficient to get all the results shown in Appendices \ref{a10} and \ref{a50}. 
We do not show the explicit expressions of the initial values in this paper because they are too complicated. 
We have also confirmed that the initial value $\tilde{\Theta}_{0,k_{\mathrm{1}},k_{\mathrm{2}}}$ becomes zero for $k_{\mathrm{1}}\ne k_{\mathrm{2}}$. 
Note that $\Theta_{0,k_{\mathrm{1}},k_{\mathrm{2}}}$
is obtained by
$\nu_{0,k_{\mathrm{1}},k_{\mathrm{2}}}\tilde{\Theta}_{0,k_{\mathrm{1}},k_{\mathrm{2}}}$
from eq.(\ref{be401}). 

\subsubsection{Evaluation of $\Xi_{k,k_{\mathrm{1}},k_{\mathrm{2}}}^{Y,r,n_{\mathrm{1}},n_{\mathrm{2}}}(\chi)$}
Using the inductive equation (\ref{be320}) and the initial value $\tilde{\Theta}_{0,k_{\mathrm{1}},k_{\mathrm{1}}}$
calculated in Sec.\ref{init}, 
the values of the integral 
$\Theta_{k,k_{\mathrm{1}},k_{\mathrm{2}}}$
for any $k$, $k_{\mathrm{1}}$ and $k_{\mathrm{2}}$, can be obtained with the relation (\ref{be401}). 
The result is that
$\Theta_{k,k_{\mathrm{1}},k_{\mathrm{2}}}$
vanishes $k=|k_{\mathrm{1}}-k_{\mathrm{2}}|+2q$, where $q$ is a positive integer or
zero. 
In order to obtain $\Xi_{k,
k_{\mathrm{1}},k_{\mathrm{2}}}^{Y,r,n_{\mathrm{1}},
n_{\mathrm{2}}}(\chi)$, it is sufficient
to have
$\Theta_{k,k_{\mathrm{1}},k_{\mathrm{2}}}$ 
only for $k_{\mathrm{1}}\ge k_{\mathrm{2}}$.  
This is because the value of $\Xi_{k, k_{\mathrm{1}},k_{\mathrm{2}}}^{Y,r,n_{\mathrm{1}}, n_{\mathrm{2}}}(\chi)$, with $k_{\mathrm{1}}$ and $k_{\mathrm{2}}$, 
$n_{\mathrm{1}}$ and $n_{\mathrm{2}}$ interchanged, corresponds to the
value of $\Xi_{k, k_{\mathrm{1}},k_{\mathrm{2}}}^{Y,r,
n_{\mathrm{1}}, n_{\mathrm{2}}}(\chi)$
with $\chi$ replaced by $\pi-\chi$. 
Thus, if $k_{\mathrm{1}}\ne k_{\mathrm{2}}$ or
$n_{\mathrm{1}}\ne n_{\mathrm{2}}$, 
$\Xi_{k,
k_{\mathrm{1}},k_{\mathrm{2}}}^{Y,r, n_{\mathrm{1}}, n_{\mathrm{2}}}(\chi)$
for any set of $k_{\mathrm{1}}$ and $k_{\mathrm{2}}$, $n_{\mathrm{1}}$
and $n_{\mathrm{2}}$ corresponds to 
\begin{eqnarray}
\Xi_{k, k_{\mathrm{1}},k_{\mathrm{2}}}^{Y,r,n_{\mathrm{1}}, n_{\mathrm{2}}}(\chi)
=\Xi_{k, k_{\mathrm{1}},k_{\mathrm{2}}}^{Y,r,n_{\mathrm{1}}, n_{\mathrm{2}}}(\chi)+\Xi_{k, k_{\mathrm{1}}, k_{\mathrm{2}}}^{Y,r,n_{\mathrm{1}}, n_{\mathrm{2}}}(\pi-\chi),  
\label{be34.2}
\end{eqnarray}
for $k_{\mathrm{1}}\ge k_{\mathrm{2}}$
; if $k_{\mathrm{1}}=k_{\mathrm{2}}$ and
$n_{\mathrm{1}}=n_{\mathrm{2}}$, then 
$\Xi_{k, k_{\mathrm{1}},k_{\mathrm{2}}}^{Y,r,n_{\mathrm{1}},
n_{\mathrm{2}}}(\chi)$ gives the required
value at once.

\section{Calculation of the First Order Coefficients $B_{kr}^{\mathrm{I}}$}\label{a10}
Let us explain how to obtain the first-order coefficients, that is, 
how to solve the integral equation (\ref{be4}). 
To begin with, we calculate $\Omega_{kr}^{\mathrm{H}}$ in
eq.(\ref{be25.5}) to first order; 
$\Omega_{kr}^{\mathrm{H}}$ for first order corresponds to the right-hand side
of eq.(\ref{be4}). 
It can be calculated only by substituting
$B_{00}=1$ into the expressions of $D_{k,r}$ and $E_{k,r}$ in eqs.(\ref{be22}) and (\ref{be23}): the coefficient $B_{00}=1$
corresponds to $f_{\mathrm{1}}=f^{(0)}_{\mathrm{1}}$, and no higher-order terms appear in $\Omega^{\mathrm{H}}_{kr}$ to first order. 
It finally becomes 
\begin{eqnarray}
\Omega^{\mathrm{H}}_{1r}=-\frac{2n}{T}\left(\frac{2\kappa T}{m}\right)^{\frac{1}{2}}\frac{\partial T}{\partial x}\delta_{1,r}. 
\label{be36}
\end{eqnarray} 

Now $\Omega^{\mathrm{H}}_{kr}$ for first order is found to vanish unless
$k=1$, so that we need calculate only $\Delta^{\mathrm{H}}_{1r}$ for first order; as was
mentioned in the end of Sec.\ref{solve}, we do not need to consider the case in
which the right-hand side of eq.(\ref{be4}) becomes zero.\cite{resi} 
To derive $\Delta^{\mathrm{H}}_{1r}$ in eq.(\ref{be28}) for first order, 
we must calculate both $W_{1,k_{\mathrm{1}},k_{\mathrm{2}}}^{r,n_{\mathrm{1}},
n_{\mathrm{2}}}$ and
$\Xi_{1,k_{\mathrm{1}},k_{\mathrm{2}}}^{r,n_{\mathrm{1}},n_{\mathrm{2}}}$ 
in $F_{1,r}^{1}(\chi)$ of eq.(\ref{be34.5}) for first order, as was shown
in Appendices \ref{a2} and \ref{kimref}.    
The result for $\Delta^{\mathrm{H}}_{1r}$ to first order can be written
finally in the form 
\begin{eqnarray}
\Delta^{\mathrm{H}}_{1r}=B_{00}\sum_{n_{\mathrm{1}}} B_{1n_{\mathrm{1}}}^{\mathrm{I}} M_{1,1,0}^{Y,r,n_{\mathrm{1}},0}, \label{be35}
\end{eqnarray}
where the set of the coefficients $B_{1n_{\mathrm{1}}}^{\mathrm{I}}B_{00}$ is obtained from 
$W_{1,1,0}^{r,n_{\mathrm{1}},0}$ in eq.(\ref{be34}). 
To first order, $f_{\mathrm{1}}$ in eq.(\ref{be27}) contains  only $B_{00}=1$ and
the first-order coefficients $B_{k_{\mathrm{1}}n_{\mathrm{1}}}^{\mathrm{I}}$ and $C_{k_{\mathrm{1}}n_{\mathrm{1}}}^{\mathrm{I}}$; $f_{\mathrm{2}}$ in
eq.(\ref{be27}) also contains $B_{00}=1$, $B_{k_{\mathrm{2}}n_{\mathrm{2}}}^{\mathrm{I}}$ and $C_{k_{\mathrm{2}}n_{\mathrm{2}}}^{\mathrm{I}}$ to first order. 
Thus, we obtain only the term $B_{1n_{\mathrm{1}}}^{\mathrm{I}}B_{00}$ from $W_{1,1,0}^{r,n_{\mathrm{1}},0}$ to first order using the
fact that $F_{kr}^{1}(\chi)=0$ unless
$k=|k_{\mathrm{1}}-k_{\mathrm{2}}|+2q$. 
Note that it is sufficient to consider only the case for
$k_{\mathrm{1}}\ge k_{\mathrm{2}}$ as is explained in Appendix \ref{kimref}, and that we set $q=0$. 
The matrix $M_{1,1,0}^{Y,r,n_{\mathrm{1}},0}$ is thus obtained 
\begin{eqnarray}
M_{1,1,0}^{Y,r,n_{\mathrm{1}},0}&=&\frac{d m n^{2}}{\pi^{2}\kappa T} \frac{n_{\mathrm{1}}!}{\Gamma(r+2)}\int_{0}^{2\pi} [\Xi_{1,1,0}^{Y,r,n_{\mathrm{1}},0}(\chi)-\Xi_{1,1,0}^{Y,r,n_{\mathrm{1}},0}(0)]\sin\frac{\chi}{2} {\mathrm d}\chi, \label{be28.1}
\end{eqnarray}
using eqs.(\ref{be28}), (\ref{be34.5}) and (\ref{be34}). 

For $k=1$, eq.(\ref{be0}) gives a simultaneous equation determining
the first-order coefficients $B_{1n_{\mathrm{1}}}^{\mathrm{I}}$, \textit{i.e.} 
\begin{eqnarray}
\Omega^{\mathrm{H}}_{1r}=\sum_{n_{\mathrm{1}}\ge 1} B_{1n_{\mathrm{1}}}^{\mathrm{I}}M_{1,1,0}^{Y,r,n_{\mathrm{1}},0},\label{be0.1}
\end{eqnarray}
from eqs.(\ref{be36}) and (\ref{be35}). 
Note that we need only to obtain the first-order coefficients
$B_{1n_{\mathrm{1}}}^{\mathrm{I}}$ for $n_{\mathrm{1}}\ge 1$, because $B_{10}=0$ from
eq.(\ref{be18}). 
We have calculated the matrix $M_{1,1,0}^{Y,r,n_{\mathrm{1}},
0}$ for $1 \le r \le 7$ and $1 \le n_{\mathrm{1}} \le 7$   from
eq.(\ref{be28.1}), and we have also confirmed that $M_{1,1,0}^{Y,0,n_{\mathrm{1}},
0}$ for $1\le n_{\mathrm{1}}\le 7$ calculated from
eq.(\ref{be28.1}) vanishes.  
Our result for $M_{1,1,0}^{Y,r,n_{\mathrm{1}},
0}$ for $1 \le r \le 7$ and $1 \le n_{\mathrm{1}} \le 7$ is given in
Appendix \ref{matrix}. 
At last, we can determine the first-order coefficients
$B_{1n_{\mathrm{1}}}^{\mathrm{I}}$ by solving the simultaneous
equation (\ref{be0.1}), that is, $B_{1n_{\mathrm{1}}}^{\mathrm{I}}$ can be obtained as
\begin{eqnarray}
B_{1n_{\mathrm{1}}}^{\mathrm{I}}=\sum_{r\ge 1} \Omega^{\mathrm{H}}_{1r} (M_{1,1,0}^{Y,r,n_{\mathrm{1}},0})^{-1}, \label{be35.5}
\end{eqnarray}
where $X^{-1}$ represents the inverse matrix of a matrix $X$. 
Finally, the results of the first-order coefficients
$B_{k_{\mathrm{1}}n_{\mathrm{1}}}^{\mathrm{I}}$, 
\textit{i.e.} the first-order $B_{kr}$ in eq.(\ref{be14}), can be 
calculated as in eq.(\ref{be37}). 

\section{Calculation of the Second Order Coefficients $B_{kr}^{\mathrm{II}}$}\label{a50} 
We explain how to obtain the second-order coefficients, that is, 
how to solve the integral equation (\ref{be5}). 
The coefficients of first order, \textit{i.e.} $B_{kr}^{\mathrm{I}}$ and $C_{kr}^{\mathrm{I}}$, have been obtained as are given in eq.(\ref{be37}), 
so that we can employ them to determine the second-order coefficients. 

To begin with, we calculate $\Omega^{\mathrm{H}}_{kr}$ in
eq.(\ref{be25.5}) for second order; $\Omega_{kr}^{\mathrm{H}}$ for
second order corresponds to the first term on the right-hand side
of eq.(\ref{be5}). 
It can be calculated by substituting $B_{kr}^{\mathrm{I}}$ and $C_{kr}^{\mathrm{I}}$ into the expressions of $D_{k,r}$ and $E_{k,r}$ in eqs.(\ref{be22}) and (\ref{be23}); no other terms appear in $\Omega^{\mathrm{H}}_{kr}$ for second order. 
The results of the tedious calculation of $\Omega^{\mathrm{H}}_{kr}$ to second order finally become as follows. 
For $k=0$, $\Omega^{\mathrm{H}}_{0r}$ becomes 
\begin{eqnarray}
\Omega^{\mathrm{H}}_{0r}=0,
\label{be41r01}
\end{eqnarray}
for $r=0$ and $1$,  
\begin{eqnarray}
\Omega^{\mathrm{H}}_{02}=-\frac{2\pi^{2}}{d T^{2}}\left(\frac{2\kappa T}{m}\right)^{\frac{1}{2}}\left\{
(\nabla T)^{2}\left(\frac{9}{2}b_{12}-\frac{7}{2}b_{11}\right)
+T(\nabla^{2} T)\left(3 b_{12}-b_{11}\right)
\right\}, \nonumber \\
\label{be41r2}
\end{eqnarray}
for $r=2$, and  
\begin{eqnarray}
\Omega^{\mathrm{H}}_{0r}&=&-\frac{2\pi^{2}}{d T^{2}}\left(\frac{2\kappa T}{m}\right)^{\frac{1}{2}} \left\{T(\nabla^{2} T)\left[(r+1)b_{1r}-b_{1,r-1}\right] \right. \nonumber \\
&+& \left. (\nabla T)^{2}\left[(r^{2}+\frac{1}{2}r-\frac{1}{2})b_{1r}-(2r-\frac{1}{2}) b_{1,r-1}+b_{1,r-2}\right]\right\}, 
\label{be41}
\end{eqnarray}
for $r \ge 3$. 
Note that values for the constants $b_{1r}$ are summarized in
Table \ref{b1r}. 
For $k=2$, $\Omega^{\mathrm{H}}_{2r}$ becomes 
\begin{eqnarray}
\Omega^{\mathrm{H}}_{20}=\frac{\pi^{2}}{d T^{2}}\left(\frac{2\kappa T}{m}\right)^{\frac{1}{2}} \left\{ \frac{b_{11}}{2}\left[\left(\partial_{x} T\right)^{2}-\left(\partial_{y} T\right)^{2}\right]+ b_{11} T\left[\partial_{x}^{2} T-\partial_{y}^{2} T\right]\right\},
\label{be44r0}
\end{eqnarray} 
and 
\begin{eqnarray}
\Omega^{\mathrm{H}}_{21}&=&-\frac{\pi^{2}}{d T^{2}}\left(\frac{2\kappa T}{m}\right)^{\frac{1}{2}}\left\{\left[\left(\partial_{x} T\right)^{2}-\left(\partial_{y} T\right)^{2}\right] \left[\frac{5}{2}b_{11}-3 b_{12}\right]+T\left[\partial_{x}^{2} T-\partial_{y}^{2} T\right] \left[b_{11}-2 b_{12}\right]\right\}, \nonumber \\
\label{be44r1}
\end{eqnarray} 
and
\begin{eqnarray}
\Omega^{\mathrm{H}}_{2r}&=&-\frac{\pi^{2}}{d T^{2}}\left(\frac{2\kappa T}{m}\right)^{\frac{1}{2}} \nonumber \\
&\times& \left\{
\left[\left(\partial_{x} T\right)^{2}-\left(\partial_{y} T\right)^{2}\right]\left[-(r+\frac{1}{2})(r+1)b_{1,r+1}+(2r+\frac{1}{2})b_{1r}-b_{1,r-1}\right] \right. \nonumber \\
&+& \left. T\left[\partial_{x}^{2} T-\partial_{y}^{2} T\right] \left[-(r+1) b_{1,r+1}+b_{1r}\right]
\right\},
\label{be44}
\end{eqnarray} 
for $r \ge 2$. 
For $k=1$ and $k\ge 3$, we find that $\Omega^{\mathrm{H}}_{kr}$ for second order becomes  
\begin{eqnarray}
\Omega^{\mathrm{H}}_{kr}=0,  
\label{be44.1}
\end{eqnarray} 
for any value of $r$. 

Next let us calculate $\Delta^{\mathrm{H}}_{kr}$ in eq.(\ref{be28}) for
second order. 
In order to derive $\Delta^{\mathrm{H}}_{kr}$ for second order, 
we have to calculate $W_{k,k_{\mathrm{1}},k_{\mathrm{2}}}^{r,n_{\mathrm{1}},
n_{\mathrm{2}}}$ and
$\Xi_{k,k_{\mathrm{1}},k_{\mathrm{2}}}^{Y,r,n_{\mathrm{1}},n_{\mathrm{2}}}$ 
in $F_{kr}^{1}(\chi)$ of eq.(\ref{be34.5}) to second order, as was shown
in Appendix \ref{a2}.    
For $k=0$, $\Delta^{\mathrm{H}}_{kr}$ to second order
results in
\begin{eqnarray}
\Delta^{\mathrm{H}}_{0r}&=&
B_{00}\sum_{n_{\mathrm{1}}\ge 2} B_{0n_{\mathrm{1}}}^{\mathrm{II}} M_{0,0,0}^{Y,r,n_{\mathrm{1}},
0}+\sum_{n_{\mathrm{1}},n_{\mathrm{2}}} B_{1n_{\mathrm{1}}}^{\mathrm{I}}B_{1n_{\mathrm{2}}}^{\mathrm{I}}M_{0,1,1}^{Y,r,n_{\mathrm{1}},
n_{\mathrm{2}}}+\sum_{n_{\mathrm{1}},n_{\mathrm{2}}} C^{\mathrm{I}}_{1n_{\mathrm{1}}}C^{\mathrm{I}}_{1n_{\mathrm{2}}} M_{0,1,1}^{Z,r,n_{\mathrm{1}},
n_{\mathrm{2}}}.   
\label{be990}
\end{eqnarray}
$B_{1n_{\mathrm{1}}}^{\mathrm{I}}$ from $f_{\mathrm{1}}$ and $B_{1n_{\mathrm{2}}}^{\mathrm{I}}$
from $f_{\mathrm{2}}$ of 
the set of the coefficients $B_{1n_{\mathrm{1}}}^{\mathrm{I}}B_{1n_{\mathrm{2}}}^{\mathrm{I}}$ 
are the first-order coefficients obtained in eq.(\ref{be37}), so that  $B_{1n_{\mathrm{1}}}^{\mathrm{I}}B_{1n_{\mathrm{2}}}^{\mathrm{I}}$ is second order. 
Similarly, $C^{\mathrm{I}}_{1n_{\mathrm{1}}}C^{\mathrm{I}}_{1n_{\mathrm{2}}}$
is also second order. 
The second and the third terms on the right-hand side of eq.(\ref{be990}) correspond to
$J(f_{\mathrm{1}},f_{\mathrm{2}})$ in the integral equation
(\ref{be5}). 
To second order, $f_{\mathrm{i}}$ of eq.(\ref{be27}) contains only $B_{00}=1$, $B_{k_{\mathrm{i}}n_{\mathrm{i}}}^{\mathrm{I}}$ and $C_{k_{\mathrm{i}}n_{\mathrm{i}}}^{\mathrm{I}}$ obtained in eq.(\ref{be37}), and $B_{k_{\mathrm{i}}n_{\mathrm{i}}}^{\mathrm{II}}$ to be determined
here for ${\mathrm i}=1$ and $2$. 
Therefore, we can only obtain the sets of the terms in eq.(\ref{be990}) for
second order by using the
fact that $F_{kr}^{1}(\chi)=0$ unless
$k=|k_{\mathrm{1}}-k_{\mathrm{2}}|+2q$.    
We should derive the second-order coefficients
$B_{0n_{\mathrm{1}}}^{\mathrm{II}}$ only for $n_{\mathrm{1}}\ge 2$, because $B_{00}=1$ and $B_{01}=0$ from
eq.(\ref{be18}). 
Note that it is sufficient to consider only the case for
$k_{\mathrm{1}}\ge k_{\mathrm{2}}$, as is explained in Appendix \ref{kimref}, and
that  $B^{\mathrm{I}}_{1n_{\mathrm{1}}}C^{\mathrm{I}}_{1n_{\mathrm{2}}}+C^{\mathrm{I}}_{1n_{\mathrm{1}}}B^{\mathrm{I}}_{1n_{\mathrm{2}}}$ does not appear. (see Appendix \ref{a2}) 

The matrix $M_{0,0,0}^{Y,r,n_{\mathrm{1}},
0}$ in eq.(\ref{be990}) is obtained as
\begin{eqnarray}
M_{0,0,0}^{Y,r,n_{\mathrm{1}},0}
&=&\frac{2 d n^{2}}{\pi^{2}} \int_{0}^{2\pi} [\Xi_{0,0,0}^{Y,r,n_{\mathrm{1}},0}(\chi)-\Xi_{0,0,0}^{Y,r,n_{\mathrm{1}},0}(0) ]\sin\frac{\chi}{2}  {\mathrm d}\chi, \label{be28.10}
\end{eqnarray}
using eqs.(\ref{be28}), (\ref{be34.5}) and (\ref{be34}). 
Similarly, the matrices $M_{0,1,1}^{Y,r,n_{\mathrm{1}},
n_{\mathrm{2}}}$ in eq.(\ref{be990}) are derived as 
\begin{eqnarray}
M_{0,1,1}^{Y,r,n_{\mathrm{1}},
n_{\mathrm{2}}}
&=&\frac{d m n^{2}}{\pi^{2} \kappa T}\int_{0}^{2\pi} [\Xi_{0,1,1}^{Y,r,n_{\mathrm{1}},n_{\mathrm{2}}}(\chi)-\Xi_{0,1,1}^{Y,r,n_{\mathrm{1}},n_{\mathrm{2}}}(0) ]\sin\frac{\chi}{2}  {\mathrm d}\chi, \nonumber \\ \label{be28.11}
\end{eqnarray}
while we have confirmed $M_{0,1,1}^{Z,r,n_{\mathrm{1}},
n_{\mathrm{2}}}=M_{0,1,1}^{Y,r,n_{\mathrm{1}},
n_{\mathrm{2}}}$. 
Equations (\ref{be990}), (\ref{be41r01}), (\ref{be41r2}) and (\ref{be41}) lead to a simultaneous equation to determine
the second-order coefficients $B_{0n_{\mathrm{1}}}^{\mathrm{II}}$: 
\begin{eqnarray}
B_{0n_{\mathrm{1}}}^{\mathrm{II}}=\sum_{r\ge 2} \left\{\Omega^{\mathrm{H}}_{0r}-\sum_{n_{\mathrm{1}},n_{\mathrm{2}}} \left(B_{1n_{\mathrm{1}}}^{\mathrm{I}}B_{1n_{\mathrm{2}}}^{\mathrm{I}}+C^{\mathrm{I}}_{1n_{\mathrm{1}}}C^{\mathrm{I}}_{1n_{\mathrm{2}}} \right) M_{0,1,1}^{Y,r,n_{\mathrm{1}},
n_{\mathrm{2}}}\right\}(M_{0}^{Y,r,n_{\mathrm{1}},
0})^{-1}. \nonumber \\ \label{be131}
\end{eqnarray}
We have calculated the matrix $M_{0,0,0}^{Y,r,n_{\mathrm{1}},
0}$ for $2 \le r \le 6$ and $2 \le n_{\mathrm{1}} \le 6$  from
eq.(\ref{be28.10}), and also confirmed that 
$M_{0,0,0}^{Y,r,n_{\mathrm{1}},
0}$ vanishes for $r=0, 1$ and $2 \le n_{\mathrm{1}} \le 6$ or for $2 \le r \le 6$ and $n_{\mathrm{1}}=0, 1$. 
We have calculated the matrix $M_{0,
1,1}^{Y,r,n_{\mathrm{1}},n_{\mathrm{2}}}$ for $2\le r \le 6$, $1\le
n_{\mathrm{1}} \le 7$ and $1\le n_{\mathrm{2}} \le 7$ from
eq.(\ref{be28.11}), and confirmed
$M_{0,
1,1}^{Y,r,n_{\mathrm{1}},n_{\mathrm{2}}}$ vanishes for $r=0,
1$, $1\le n_{\mathrm{1}} \le 7$ and $1\le n_{\mathrm{2}} \le 7$. 
Our results for $M_{0,0,0}^{Y,r,n_{\mathrm{1}},
0}$ for $2 \le r \le 6$ and $2 \le n_{\mathrm{1}} \le 6$  and $M_{0,
1,1}^{Y,r,n_{\mathrm{1}},n_{\mathrm{2}}}$ for $2\le r \le 6$, $1\le n_{\mathrm{1}}
\le 7$ and $1\le n_{\mathrm{2}} \le 7$ are given in Appendix \ref{matrix}. 
Finally, we can determine the second-order coefficients
$B_{0n_{\mathrm{1}}}^{\mathrm{II}}$ in $f_{\mathrm{1}}$,
\textit{i.e.} the second-order $B_{0r}$ in eq.(\ref{be13}) as in eq.(\ref{be42AB}). 

Similarly, for $k=2$, $\Delta^{\mathrm{H}}_{kr}$ for second order
results in
\begin{eqnarray}
\Delta^{\mathrm{H}}_{2r}&=&
B_{00}\sum_{n_{\mathrm{1}}\ge 0} B_{2n_{\mathrm{1}}}^{\mathrm{II}} M_{2,2,0}^{Y,r,n_{\mathrm{1}},
0}+\sum_{n_{\mathrm{1}},n_{\mathrm{2}}} B_{1n_{\mathrm{1}}}^{\mathrm{I}}B_{1n_{\mathrm{2}}}^{\mathrm{I}}M_{2,1,1}^{Y,r,n_{\mathrm{1}},
n_{\mathrm{2}}}+\sum_{n_{\mathrm{1}},n_{\mathrm{2}}} C^{\mathrm{I}}_{1n_{\mathrm{1}}}C^{\mathrm{I}}_{1n_{\mathrm{2}}} M_{2,1,1}^{Z,r,n_{\mathrm{1}},
n_{\mathrm{2}}},   
\label{be991}
\end{eqnarray}
using the fact that $F_{kr}^{1}(\chi)=0$ unless
$k=|k_{\mathrm{1}}-k_{\mathrm{2}}|+2q$.    
Note that we have confirmed $M_{2,0,0}^{Y,r,n_{\mathrm{1}},
0}$ becomes zero. 
The matrix $M_{2,2,0}^{Y,r,n_{\mathrm{1}},
0}$ in eq.(\ref{be991}) is obtained as
\begin{eqnarray}
M_{2,2,0}^{Y,r,n_{\mathrm{1}},0}
&=&\frac{d m^{2} n^{2}}{2 \pi^{2} \kappa^{2} T^{2}} \int_{0}^{2\pi} [\Xi_{2,2,0}^{Y,r,n_{\mathrm{1}},0}(\chi)-\Xi_{2,2,0}^{Y,r,n_{\mathrm{1}},0}(0) ]\sin\frac{\chi}{2}  {\mathrm d}\chi, \label{be28.100}
\end{eqnarray}
using eqs.(\ref{be28}), (\ref{be34.5}) and (\ref{be34}). 
Similarly, the matrices $M_{2,1,1}^{Y,r,n_{\mathrm{1}},
n_{\mathrm{2}}}$ in eq.(\ref{be991}) are derived as 
\begin{eqnarray}
M_{2,1,1}^{Y,r,n_{\mathrm{1}},
n_{\mathrm{2}}}
&=&\frac{d m^{2} n^{2}}{2 \pi^{2} \kappa^{2} T^{2}}\int_{0}^{2\pi} [\Xi_{2,1,1}^{Y,r,n_{\mathrm{1}},n_{\mathrm{2}}}(\chi)-\Xi_{2,1,1}^{Y,r,n_{\mathrm{1}},n_{\mathrm{2}}}(0) ]\sin\frac{\chi}{2}  {\mathrm d}\chi, \nonumber \\ \label{be28.111}
\end{eqnarray}
while we have confirmed $M_{2,1,1}^{Z,r,n_{\mathrm{1}},
n_{\mathrm{2}}}=-M_{2,1,1}^{Y,r,n_{\mathrm{1}},
n_{\mathrm{2}}}$. 
Thus, eqs. (\ref{be991}), (\ref{be44r0}), (\ref{be44r1}) and (\ref{be44}) lead to a simultaneous equation to determine
the second-order coefficients $B_{2n_{\mathrm{1}}}^{\mathrm{II}}$: 
\begin{eqnarray}
B_{2n_{\mathrm{1}}}^{\mathrm{II}}=\sum_{r\ge 0} \left\{\Omega^{\mathrm{H}}_{2r}-\sum_{n_{\mathrm{1}},n_{\mathrm{2}}} (B_{1n_{\mathrm{1}}}^{\mathrm{I}}B_{1n_{\mathrm{2}}}^{\mathrm{I}}-C^{\mathrm{I}}_{1n_{\mathrm{1}}}C^{\mathrm{I}}_{1n_{\mathrm{2}}}) M_{2,1,1}^{Y,r,n_{\mathrm{1}},
n_{\mathrm{2}}}\right\}(M_{2,2,0}^{Y,r,n_{\mathrm{1}},
0})^{-1}. \label{be123}
\end{eqnarray}
In order to derive the second-order coefficients
$B_{2n_{\mathrm{1}}}^{\mathrm{II}}$ for $n_{\mathrm{1}}\ge 0$,  
we have calculated the matrix $M_{2,2,0}^{Y,r,n_{\mathrm{1}},
0}$ for $0\le r \le 6$ and $0 \le n_{\mathrm{1}} \le 6$  from
eq.(\ref{be28.100}), and also the matrices $M_{2,
1,1}^{Y,r,n_{\mathrm{1}},n_{\mathrm{2}}}$ for $0\le r
\le 6$ , $1\le n_{\mathrm{1}}\le 7$ and $1\le n_{\mathrm{2}}\le 7$ from eq.(\ref{be28.111}). 
Those results are given in Appendix \ref{matrix}. 
The second-order coefficients
$B_{k_{\mathrm{1}}n_{\mathrm{1}}}^{\mathrm{II}}$ in $f_{\mathrm{1}}$,
\textit{i.e.} the second-order $B_{kr}$ in eq.(\ref{be14}), can be written
 in the final form shown in eq.(\ref{be45AB}). 

We need to consider eq.(\ref{be0}) only for $k=0$
and $2$ for second order:  
it is not necessary to consider eq.(\ref{be0}) for even $k$ furthermore, which was first expected in ref.\cite{kim} and recently confirmed in ref.\cite{fushiki}.  
For odd $k$, 
$\Omega^{\mathrm{H}}_{kr}$
to second order is found to be zero, and no terms corresponding to $J(f_{\mathrm{1}},f_{\mathrm{2}})$ in the integral equation
(\ref{be5}), \textit{i.e.} the second and the third terms on the right-hand side of
eqs.(\ref{be131}) or (\ref{be123}) appear, so that any second-order terms $B^{\mathrm{II}}_{kr}$ do not appear for odd $k$.\cite{kim} 

\section{Matrix Elements}\label{matrix}

\subsection{$M_{\lowercase{1,1,0}}^{Y,r,n_{\mathrm{1}},0}$ for
 $\lowercase{1\le r \le 7}$ and $\lowercase{1\le n_{\mathrm{1}}\le 7}$}

The matrix elements $M_{1,1,0}^{Y,r,n_{\mathrm{1}},0}$ for $1\le r \le 7$ and $1\le n_{\mathrm{1}}\le 7$ divided by
$M_{1,1,0}^{Y,1,1,0}=-2 d n^{2} \left(\frac{\pi \kappa
 T}{m}\right)^{\frac{1}{2}}$
calculated from eq.(\ref{be28.1}) are given as
follows. 

{\scriptsize

\[\left(
\begin{array}{@{\,}ccccccc@{\,}}
{1}&{-8.333\times 10^{-2}}&{-2.604\times 10^{-3}}&{-1.302\times 10^{-4}}&{-6.782\times 10^{-6}}&{-3.391\times 10^{-7}}&{-1.589\times 10^{-8}}\\ 
{-5.000\times 10^{-1}}&{1.625}&{-1.185\times 10^{-1}}&{-3.451\times 10^{-3}}&{-1.648\times 10^{-4}}&{-8.308\times 10^{-6}}&{-4.053\times 10^{-7}}\\ 
{-1.875\times 10^{-1}}&{-1.422}&{2.165}&{-1.327\times 10^{-1}}&{-3.466\times 10^{-3}}&{-1.531\times 10^{-4}}&{-7.279\times 10^{-6}} \\ 
{-1.875\times 10^{-1}}&{-8.281\times 10^{-1}}&{-2.654}&{2.645}&{-1.383\times 10^{-1}}&{-3.234\times 10^{-3}}&{-1.314\times 10^{-4}} \\
{-2.930\times 10^{-1}}&{-1.187}&{-2.079}&{-4.148}&{3.081}&{-1.398\times 10^{-1}}&{-2.94716687\times 10^{-3}} \\
{-6.152\times 10^{-1}}&{-2.512}&{-3.859}&{-4.074}&{-5.873}&{3.483}&{-1.395\times 10^{-1}} \\
{-1.615}&{-6.864}&{-1.027\times 10}&{-9.273}&{-6.932}&{-7.810}&{3.858} \\
\end{array}
\right)  
 \]
}

\subsection{$M_{\lowercase{0,0,0}}^{Y,r,n_{\mathrm{1}},0}$ for $\lowercase{2\le r \le 6}$ and $\lowercase{2\le n_{\mathrm{1}}\le 6}$, and
 $M_{\lowercase{0,1,1}}^{Y,r,n_{\mathrm{1}},n_{\mathrm{2}}}$ for $\lowercase{2\le r \le 6}$, $\lowercase{1\le n_{\mathrm{1}} \le 7}$ and $\lowercase{1 \le n_{\mathrm{2}} \le 7}$}

The matrix elements $M_{0,0,0}^{Y,r,n_{\mathrm{1}},
0}$ for $2\le r \le 6$ and $2\le n_{\mathrm{1}}\le 6$ divided by
$M_{0,0,0}^{Y,2,2,0}=-4 d n^{2}\left(\frac{\pi \kappa
T}{m}\right)^{\frac{1}{2}}$ calculated from
eq.(\ref{be28.10}) are given as

{\footnotesize
\[\left(
\begin{array}{@{\,}ccccc@{\,}}
{1}&{-8.333\times 10^{-2}}&{-2.604\times 10^{-3}}&{-1.302\times 10^{-4}}&{-6.782\times 10^{-1}}\\ 
{-7.500\times 10^{-1}}&{1.688}&{-1.152\times 10^{-1}}&{-3.223\times 10^{-3}}&{-1.495\times 10^{-4}}\\ 
{-3.750\times 10^{-1}}&{-1.844}&{2.251}&{-1.287\times 10^{-1}}&{-3.204\times 10^{-3}} \\
{-4.688\times 10^{-1}}&{-1.289}&{-3.217}&{2.741}&{-1.342\times 10^{-1}} \\
{-8.789\times 10^{-1}}&{-2.153}&{-2.883}&{-4.833}&{3.182} \\
\end{array}
\right)  
 \]
}

The matrix elements $M_{0,1,1}^{Y,2,n_{\mathrm{1}},n_{\mathrm{2}}}$ for $1\le
 n_{\mathrm{1}} \le 7$ and $1 \le n_{\mathrm{2}} \le 7$ divided by
$M_{0,1,1}^{Y,2,1,1}=-\frac{\pi^{4} (\nabla T)^{2}}{4 d T^{2}} \left(\frac{\pi \kappa
 T}{m}\right)^{\frac{1}{2}}$ calculated from
 eq.(\ref{be28.11}) are given as

{\scriptsize
\[\left(
\begin{array}{@{\,}ccccccc@{\,}}
{1}&{5.000\times 10^{-1}}&{0.000}&{-9.375\times 10^{-1}}&{-4.102}&{-1.661\times 10}&{-7.106\times 10}\\ 
{5.000\times 10^{-1}}&{3.750\times 10^{-1}}&{9.375\times 10^{-1}}&{8.203\times 10^{-1}}&{-1.846}&{-2.030\times 10}&{-1.320\times 10^{2}}\\ 
{0.000}&{9.375\times 10^{-1}}&{1.230}&{5.537}&{1.015\times 10}&{0.000}&{-1.856\times 10^{2}}\\ 
{-9.375\times 10^{-1}}&{8.203\times 10^{-1}}&{5.537}&{1.015\times 10}&{6.598\times 10}&{1.856\times 10^{2}}&{2.629\times 10^{2}}\\ 
{-4.102}&{-1.846}&{1.015\times 10}&{6.598\times 10}&{1.547\times 10^{2}}&{1.315\times 10^{3}}&{4.995\times 10^{3}}\\ 
{-1.661\times 10}&{-2.030\times 10^{1}}&{0.000}&{1.856\times 10^{2}}&{1.315\times 10^{3}}&{3.746\times 10^{3}}&{3.934\times 10^{4}}\\ 
{-7.106\times 10}&{-1.320\times 10^{2}}&{-1.856\times 10^{2}}&{2.629\times 10^{2}}&{4.995\times 10^{3}}&{3.934\times 10^{4}}&{1.319\times 10^{5}}\\ 
\end{array}
\right)  
 \]
}

The matrix elements $M_{0,1,1}^{Y,3,n_{\mathrm{1}},n_{\mathrm{2}} }$ for $1\le
 n_{\mathrm{1}} \le 7$ and $1 \le n_{\mathrm{2}} \le 7$ divided by
$M_{0,1,1}^{Y,2,1,1}$ are given as

{\scriptsize
\[\left(
\begin{array}{@{\,}ccccccc@{\,}}
{-2.500\times 10^{-1}}&{1.125}&{5.000\times 10^{-1}}&{-4.688\times 10^{-2}}&{-2.490}&{-1.482\times 10}&{-7.913\times 10}\\ 
{1.125}&{2.813\times 10^{-1}}&{7.969\times 10^{-1}}&{1.436}&{2.000}&{-4.153}&{-7.360\times 10}\\ 
{5.000\times 10^{-1}}&{7.969\times 10^{-1}}&{7.471\times 10^{-1}}&{3.845}&{1.130\times 10^{1}}&{3.045\times 10^{1}}&{2.990\times 10^{1}}\\ 
{-4.688\times 10^{-2}}&{1.436}&{3.845}&{5.768}&{4.187\times 10}&{1.681\times 10^{2}}&{6.534\times 10^{2}}\\ 
{-2.490}&{2.000}&{1.130\times 10}&{4.187\times 10}&{8.506\times 10}&{7.926\times 10^{2}}&{4.009\times 10^{3}}\\ 
{-1.482\times 10}&{-4.153}&{3.045\times 10}&{1.681\times 10^{2}}&{7.926\times 10^{2}}&{2.021\times 10^{3}}&{2.295\times 10^{4}}\\ 
{-7.913\times 10}&{-7.360\times 10}&{2.990\times 10}&{6.534\times 10^{2}}&{4.009\times 10^{3}}&{2.295\times 10^{4}}&{7.027\times 10^{4}}\\ 
\end{array}
\right)  
 \]
}

The matrix elements $M_{0,1,1}^{Y,4,n_{\mathrm{1}},n_{\mathrm{2}}}$ for $1\le
 n_{\mathrm{1}} \le 7$ and $1 \le n_{\mathrm{2}} \le 7$ divided by
$M_{0,1,1}^{Y,2,1,1}$ are given as

{\scriptsize
\[\left(
\begin{array}{@{\,}ccccccc@{\,}}
{-1.302\times 10^{-2}}&{-2.305\times 10^{-1}}&{7.813\times 10^{-1}}&{4.946\times 10^{-1}}&{-8.331\times 10^{-2}}&{-4.567}&{-3.443\times 10}\\ 
{-2.305\times 10^{-1}}&{2.295\times 10^{-1}}&{3.960\times 10^{-1}}&{7.784\times 10^{-1}}&{1.869}&{3.623}&{-6.020}\\ 
{7.813\times 10^{-1}}&{3.960\times 10^{-1}}&{3.062\times 10^{-1}}&{1.513}&{4.918}&{1.817\times 10}&{6.396\times 10}\\ 
{4.946\times 10^{-1}}&{7.784\times 10^{-1}}&{1.513}&{2.111}&{1.527\times 10}&{6.642\times 10}&{3.216\times 10^{2}}\\ 
{-8.331\times 10^{-2}}&{1.869}&{4.918}&{1.527\times 10}&{2.971\times 10}&{2.782\times 10^{2}}&{1.502\times 10^{3}}\\ 
{-4.567}&{3.623}&{1.817\times 10}&{6.642\times 10}&{2.782\times 10^{2}}&{6.882\times 10^{2}}&{7.871\times 10^{3}}\\ 
{-3.443\times 10}&{-6.020}&{6.396\times 10^{1}}&{3.216\times 10^{2}}&{1.502\times 10^{3}}&{7.871\times 10^{3}}&{2.355\times 10^{4}}\\ 
\end{array}
\right)  
 \]
}

The matrix elements $M_{0,1,1}^{Y,5,n_{\mathrm{1}},n_{\mathrm{2}}}$ for $1\le
 n_{\mathrm{1}} \le 7$ and $1 \le n_{\mathrm{2}} \le 7$ divided by
$M_{0,1,1}^{Y,2,1,1}$ are given as

{\scriptsize
\[\left(
\begin{array}{@{\,}ccccccc@{\,}}
{-9.115\times 10^{-4}}&{-1.074\times 10^{-3}}&{-1.289\times 10^{-1}}&{5.948\times 10^{-1}}&{4.914\times 10^{-1}}&{-1.037\times 10^{-1}}&{-7.105}\\ 
{-1.074\times 10^{-2}}&{-4.272\times 10^{-2}}&{2.509\times 10^{-1}}&{3.006\times 10^{-1}}&{7.575\times 10^{-1}}&{2.279}&{5.713}\\ 
{-1.289\times 10^{-1}}&{2.509\times 10^{-1}}&{1.104\times 10^{-1}}&{4.718\times 10^{-1}}&{1.484}&{5.886}&{2.621\times 10}\\ 
{5.948\times 10^{-1}}&{3.006\times 10^{-1}}&{4.718\times 10^{-1}}&{6.042\times 10^{-1}}&{4.226}&{1.845}&{9.461\times 10}\\ 
{4.914\times 10^{-1}}&{7.575\times 10^{-1}}&{1.484}&{4.226}&{7.918}&{7.326\times 10}&{4.000\times 10^{2}}\\ 
{-1.037\times 10^{-1}}&{2.279}&{5.886}&{1.845\times 10}&{7.326\times 10}&{1.772\times 10^{2}}&{2.017\times 10}\\ 
{-7.105}&{5.713}&{2.621\times 10}&{9.461\times 10^{2}}&{4.000\times 10}&{2.017\times 10^{3}}&{5.941\times 10^{3}}\\ 
\end{array}
\right)  
 \]
}

The matrix elements $M_{0,1,1}^{Y,6,n_{\mathrm{1}},n_{\mathrm{2}} }$ for $1\le
 n_{\mathrm{1}} \le 7$ and $1 \le n_{\mathrm{2}} \le 7$ divided by
$M_{0,1,1}^{Y,2,1,1}$ are given as

{\scriptsize
\[\left(
\begin{array}{@{\,}ccccccc@{\,}}
{-6.104\times 10^{-5}}&{-6.978\times 10^{-4}}&{-5.046\times 10^{-3}}&{-8.158\times 10^{-2}}&{4.782\times 10^{-1}}&{4.898\times 10^{-1}}&{-1.087\times 10^{-1}}\\ 
{-6.978\times 10^{-4}}&{-1.976\times 10^{-3}}&{-3.976\times 10^{-2}}&{1.575\times 10^{-1}}&{2.407\times 10^{-1}}&{7.397\times 10^{-1}}&{2.678}\\ 
{-5.046\times 10^{-3}}&{-3.976\times 10^{-2}}&{5.664\times 10^{-2}}&{1.387\times 10^{-1}}&{3.788\times 10^{-1}}&{1.450}&{6.802}\\ 
{-8.158\times 10^{-2}}&{1.575\times 10^{-1}}&{1.387\times 10^{-1}}&{1.505\times 10^{-1}}&{9.806\times 10^{-1}}&{4.159}&{2.137\times 10}\\ 
{4.782\times 10^{-1}}&{2.407\times 10^{-1}}&{3.788\times 10^{-1}}&{9.806\times 10^{-1}}&{1.749}&{1.579\times 10}&{8.557\times 10}\\ 
{4.898\times 10^{-1}}&{7.397\times 10^{-1}}&{1.450}&{4.159}&{1.579\times 10}&{3.727\times 10}&{4.194\times 10^{2}}\\ 
{-1.087\times 10^{-1}}&{2.678}&{6.802}&{2.137\times 10}&{8.557\times 10}&{4.194\times 10^{2}}&{1.217\times 10^{3}}\\ 
\end{array}
\right)  
 \]
}

\subsection{$M_{\lowercase{2,2,0}}^{Y,r,n_{\mathrm{1}},0}$ for
 $\lowercase{0\le r \le 6}$ and $\lowercase{0\le n_{\mathrm{1}}\le 6}$, and
$M_{\lowercase{2,1,1}}^{Y,r,n_{\mathrm{1}},n_{\mathrm{2}}}$ for $\lowercase{0\le r \le 6}$,
$\lowercase{1\le n_{\mathrm{1}}\le 7}$ and $1\le n_{\mathrm{2}}\le 7$}

The matrix elements $M_{2,2,0}^{Y,r,n_{\mathrm{1}},0}$ for $0\le r \le 6$ and $0\le n_{\mathrm{1}}\le 6$ divided by
$M_{2,2,0}^{Y,0,0,0}=-4 d n^{2} \left(\frac{\pi \kappa
 T}{m}\right)^{\frac{1}{2}}$ calculated from
eq.(\ref{be28.100}) are given as   

{\scriptsize
\[\left(
\begin{array}{@{\,}ccccccc@{\,}}
{1}&{-8.333\times 10^{-2}}&{-2.604\times 10^{-3}}&{-1.302\times 10^{-4}}&{-6.782\times 10^{-6}}&{-3.391\times 10^{-7}}&{-1.589\times 10^{-8}} \\ 
{-2.500\times 10^{-1}}&{1.063}&{-8.008\times 10^{-2}}&{-2.376\times 10^{-3}}&{-1.149\times 10^{-4}}&{-5.849\times 10^{-6}}&{-2.874\times 10^{-7}} \\ 
{-6.250\times 10^{-2}}&{-6.406\times 10^{-1}}&{1.232}&{-7.939\times 10^{-2}}&{-2.143\times 10^{-3}}&{-9.707\times 10^{-5}}&{-4.706\times 10^{-6}} \\ 
{-4.688\times 10^{-2}}&{-2.852\times 10^{-1}}&{-1.191}&{1.420}&{-7.825\times 10^{-2}}&{-1.900\times 10^{-3}}&{-7.953\times 10^{-5}} \\ 
{-5.859\times 10^{-2}}&{-3.311\times 10^{-1}}&{-7.713\times 10^{-1}}&{-1.878}&{1.607}&{-7.676\times 10^{-2}}&{-1.680\times 10^{-3}}\\
{-1.025\times 10^{-1}}&{-5.896\times 10^{-1}}&{-1.223}&{-1.596}&{-2.687}&{1.787}&{-7.510\times 10^{-2}}\\
{-2.307\times 10^{-1}}&{-1.391}&{-2.846}&{-3.207}&{-2.822}&{-3.605}&{1.961}\\
\end{array}
\right)  
 \]
}

The matrix elements $M_{2,1,1}^{Y,0,n_{\mathrm{1}},n_{\mathrm{2}}}$ for $1\le n_{\mathrm{1}}\le 7$ and $1\le n_{\mathrm{2}}\le 7$  divided by
$M_{2,1,1}^{Y,0,1,1}=-\frac{\pi^{4}}{16 d T^{2}} \left(\frac{\pi \kappa
 T}{m}\right)^{\frac{1}{2}}\{(\partial_{x} T)^{2}-(\partial_{y} T)^{2}\}$ calculated from
eq.(\ref{be28.111}) are given as 

{\scriptsize
\[\left(
\begin{array}{@{\,}ccccccc@{\,}}
{1}&{1.500}&{1.875}&{3.281}&{7.383}&{2.030\times 10}&{6.598\times 10} \\ 
{1.500}&{9.375\times 10^{-1}}&{3.281}&{7.383}&{2.030\times 10}&{6.598\times 10}&{2.474\times 10^{2}} \\ 
{1.875}&{3.281}&{3.691}&{2.030\times 10}&{6.598\times 10}&{2.474\times 10^{2}}&{1.052\times 10^{3}} \\ 
{3.281}&{7.383}&{2.030\times 10^{1}}&{3.299\times 10^{1}}&{2.474\times 10^{2}}&{1.052\times 10^{3}}&{4.995\times 10^{3}} \\ 
{7.383}&{2.030\times 10}&{6.598\times 10}&{2.474\times 10^{2}}&{5.258\times 10^{2}}&{4.995\times 10^{3}}&{2.622\times 10^{4}} \\
{2.030\times 10}&{6.598\times 10}&{2.474\times 10^{2}}&{1.052\times 10^{3}}&{4.995\times 10^{3}}&{1.311\times 10^{4}}&{1.508\times 10^{5}} \\
{6.598\times 10}&{2.474\times 10^{2}}&{1.052\times 10^{3}}&{4.995\times 10^{3}}&{2.622\times 10^{4}}&{1.508\times 10^{5}}&{4.712\times 10^{5}} \\
\end{array}
\right)  
 \]
}

The matrix elements $M_{2,1,1}^{Y,1,n_{\mathrm{1}},n_{\mathrm{2}}}$ for $1\le n_{\mathrm{1}}\le 7$ and $1\le n_{\mathrm{2}}\le 7$  divided by
$M_{2,1,1}^{Y,0,1,1}$ are given as 

{\scriptsize
\[\left(
\begin{array}{@{\,}ccccccc@{\,}}
{2.250}&{2.125}&{3.094}&{6.680}&{1.825\times 10}&{5.952\times 10}&{2.246\times 10^{2}} \\ 
{2.125}&{7.969\times 10^{-1}}&{2.930}&{8.408}&{2.999\times 10}&{1.231\times 10^{2}}&{5.650\times 10^{2}} \\ 
{3.094}&{2.930}&{2.563}&{1.523\times 10}&{6.218\times 10}&{3.011\times 10^{2}}&{1.624\times 10^{2}} \\ 
{6.680}&{8.408}&{15.23\times 10}&{2.094\times 10}&{1.691\times 10^{2}}&{8.815\times 10^{2}}&{5.324\times 10^{3}} \\ 
{1.825\times 10}&{2.999\times 10}&{6.218\times 10}&{1.691\times 10^{2}}&{3.170\times 10^{2}}&{3.221\times 10^{3}}&{2.029\times 10^{3}} \\
{5.952\times 10^{1}}&{1.231\times 10^{2}}&{3.011\times 10^{2}}&{8.815\times 10^{2}}&{3.221\times 10^{3}}&{7.649\times 10^{3}}&{9.343\times 10^{4}} \\
{2.246\times 10^{2}}&{5.650\times 10^{2}}&{1.624\times 10^{3}}&{5.324\times 10^{3}}&{2.029\times 10^{4}}&{9.343\times 10^{4}}&{2.686\times 10^{5}} \\
\end{array}
\right)  
 \]
}

The matrix elements $M_{2,1,1}^{Y,2,n_{\mathrm{1}},n_{\mathrm{2}}}$ for $1\le n_{\mathrm{1}}\le 7$ and $1\le n_{\mathrm{2}}\le 7$  divided by
$M_{2,1,1}^{Y,0,1,1}$ are given as 

{\scriptsize
\[\left(
\begin{array}{@{\,}ccccccc@{\,}}
{-3.211}&{3.855}&{2.370}&{4.590}&{1.368\times 10}&{5.090\times 10}&{2.199\times 10^{2}} \\ 
{3.855}&{8.413\times 10^{-1}}&{2.059}&{4.478}&{1.522\times 10}&{6.873\times 10}&{3.660\times 10^{2}} \\ 
{2.370}&{2.059}&{1.799}&{8.862}&{2.974\times 10}&{1.351\times 10^{2}}&{7.762\times 10^{2}} \\ 
{4.590}&{4.478}&{8.862}&{1.268\times 10}&{9.068\times 10}&{4.030\times 10^{2}}&{2.253\times 10^{3}} \\ 
{1.368\times 10}&{1.522\times 10}&{2.974\times 10}&{9.068\times 10}&{1.778\times 10^{2}}&{1.650\times 10^{3}}&{9.068\times 10^{3}} \\
{5.090\times 10}&{6.873\times 10}&{1.351\times 10^{2}}&{4.030\times 10^{2}}&{1.650\times 10^{3}}&{4.090\times 10^{3}}&{4.651\times 10^{4}} \\
{2.199\times 10^{2}}&{3.660\times 10^{2}}&{7.762\times 10^{2}}&{2.253\times 10^{3}}&{9.068\times 10^{3}}&{4.651\times 10^{4}}&{1.391\times 10^{5}} \\
\end{array}
\right)  
 \]
}

The matrix elements $M_{2,1,1}^{Y,3,n_{\mathrm{1}},n_{\mathrm{2}}}$ for $1\le n_{\mathrm{1}}\le 7$ and $1\le n_{\mathrm{2}}\le 7$  divided by
$M_{2,1,1}^{Y,0,1,1}$ are given as 

{\scriptsize
\[\left(
\begin{array}{@{\,}ccccccc@{\,}}
{2.493\times 10^{-1}}&{-3.585}&{3.119}&{2.456}&{5.985}&{2.271\times 10}&{1.070\times 10^{2}} \\ 
{-3.585}&{1.265}&{1.474}&{2.290}&{5.928}&{2.340\times 10}&{1.257\times 10^{2}} \\ 
{3.119}&{1.474}&{9.898\times 10^{-1}}&{4.274}&{1.257\times 10}&{4.831\times 10}&{2.443\times 10^{2}} \\ 
{2.456}&{2.290}&{4.274}&{5.626}&{3.800\times 10}&{1.565\times 10^{2}}&{7.774\times 10^{2}} \\ 
{5.985}&{5.928}&{1.257\times 10}&{3.800\times 10}&{7.156\times 10}&{6.429\times 10^{2}}&{3.357\times 10^{3}} \\
{2.271\times 10}&{2.340\times 10}&{4.831\times 10}&{1.565\times 10^{2}}&{6.429\times 10^{2}}&{1.557\times 10^{3}}&{1.733\times 10^{4}} \\
{1.070\times 10^{2}}&{1.257\times 10^{2}}&{2.443\times 10^{2}}&{7.774\times 10^{2}}&{3.357\times 10^{3}}&{1.733\times 10^{4}}&{5.108\times 10^{4}} \\
\end{array}
\right)  
 \]
}

The matrix elements $M_{2,1,1}^{Y,4,n_{\mathrm{1}},n_{\mathrm{2}}}$ for $1\le n_{\mathrm{1}}\le 7$ and $1\le n_{\mathrm{2}}\le 7$  divided by
$M_{2,1,1}^{Y,0,1,1}$ are given as 

{\scriptsize
\[\left(
\begin{array}{@{\,}ccccccc@{\,}}
{7.719\times 10^{-3}}&{2.301\times 10^{-1}}&{-2.277}&{2.579}&{2.484}&{7.307}&{3.369\times 10} \\
{2.301\times 10^{-1}}&{-8.267\times 10^{-1}}&{1.678}&{1.262}&{2.404}&{7.292}&{3.285\times 10} \\
{-2.277}&{1.678}&{5.255\times 10^{-1}}&{1.778}&{4.645}&{1.607\times 10}&{7.049\times 10} \\
{2.579}&{1.262}&{1.778}&{2.037}&{1.276\times 10}&{5.015\times 10}&{2.363\times 10^{2}} \\
{2.484}&{2.404}&{4.645}&{1.276\times 10}&{2.262\times 10}&{1.960\times 10^{2}}&{9.990\times 10^{2}} \\
{7.307}&{7.2919}&{1.607\times 10}&{5.015\times 10}&{1.960\times 10^{2}}&{4.589\times 10^{2}}&{5.000\times 10^{3}} \\
{3.369\times 10}&{3.285\times 10}&{7.049\times 10}&{2.363\times 10^{2}}&{9.990\times 10^{2}}&{5.000\times 10^{3}}&{1.442\times 10^{4}} \\
\end{array}
\right)  
 \]
}

The matrix elements $M_{2,1,1}^{Y,5,n_{\mathrm{1}},n_{\mathrm{2}}}$ for $1\le n_{\mathrm{1}}\le 7$ and $1\le n_{\mathrm{2}}\le 7$  divided by
$M_{2,1,1}^{Y,0,1,1}$ are given as 

{\scriptsize
\[\left(
\begin{array}{@{\,}ccccccc@{\,}}
{3.858\times 10^{-4}}&{6.369\times 10^{-3}}&{1.235\times 10^{-1}}&{-1.567}&{2.185}&{2.488}&{8.576} \\
{6.369\times 10^{-3}}&{4.532\times 10^{-2}}&{-8.947\times 10^{-1}}&{1.178}&{1.088}&{2.463}&{8.595} \\
{1.235\times 10^{-1}}&{-8.947\times 10^{-1}}&{4.736\times 10^{-1}}&{7.623\times 10^{-1}}&{1.573}&{4.855}&{1.940\times 10} \\
{-1.567}&{1.178}&{7.623\times 10^{-1}}&{6.743\times 10^{-1}}&{3.723}&{1.366\times 10}&{6.171\times 10} \\
{2.185}&{1.088}&{1.573}&{3.723}&{6.063}&{4.991\times 10}&{2.467\times 10^{2}} \\
{2.488}&{2.463}&{4.855}&{1.366\times 10}&{4.991\times 10}&{1.119\times 10^{2}}&{1.186\times 10^{3}} \\
{8.576}&{8.595}&{1.940\times 10}&{6.171\times 10}&{2.467\times 10^{2}}&{1.186\times 10^{3}}&{3.331\times 10^{3}} \\
\end{array}
\right)  
 \]
}

The matrix elements $M_{2,1,1}^{Y,6,n_{\mathrm{1}},n_{\mathrm{2}}}$ for $1\le n_{\mathrm{1}}\le 7$ and $1\le n_{\mathrm{2}}\le 7$  divided by
$M_{2,1,1}^{Y,0,1,1}$ are given as 

{\scriptsize
\[\left(
\begin{array}{@{\,}ccccccc@{\,}}
{2.012\times 10^{-5}}&{2.953\times 10^{-4}}&{2.997\times 10^{-3}}&{7.361\times 10^{-2}}&{-1.141}&{1.888}&{2.482} \\
{2.953\times 10^{-4}}&{1.169\times 10^{-3}}&{4.239\times 10^{-2}}&{-5.364\times 10^{-1}}&{8.671\times 10^{-1}}&{9.507\times 10^{-1}}&{2.494} \\
{2.997\times 10^{-3}}&{4.239\times 10^{-2}}&{-2.109\times 10^{-1}}&{5.791\times 10^{-1}}&{5.711\times 10^{-1}}&{1.395}&{4.977} \\
{7.361\times 10^{-2}}&{-5.364\times 10^{-1}}&{5.791\times 10^{-1}}&{2.404\times 10^{-1}}&{1.035}&{3.364}&{1.422\times 10} \\
{-1.141}&{8.671\times 10^{-1}}&{5.711\times 10^{-1}}&{1.035}&{1.473}&{1.122\times 10}&{5.291\times 10} \\
{1.888}&{9.507\times 10^{-1}}&{1.395}&{3.364}&{1.122\times 10}&{2.374\times 10}&{2.425\times 10^{2}} \\
{2.482}&{2.494}&{4.977}&{1.422\times 10}&{5.291\times 10}&{2.425\times 10^{2}}&{6.591\times 10^{2}} \\
\end{array}
\right)  
 \]
}

\newpage

\newpage

\newpage

\newpage

\newpage

\newpage

\begin{table}[htbp]
\begin{center}
\caption{\label{b1r}Numerical constants $b_{1r}$ in
 eq.(\ref{be37})}
\begin{tabular}{@{\hspace{\tabcolsep}\extracolsep{\fill}}cccccccc} \hline 
{$r$}&{$r \le 1$}&{$r \le 2$}&{$r \le 3$}&{$r \le 4$}&{$r \le 5$}&{$r \le 6$}&{$r \le 7$} \\ \hline
{$1$}&{$1$}&{$1.026$}&{$1.029$}&{$1.030$}&{$1.030$}&{$1.030$}&{$1.030$} \\ \hline
{$2$}&{$-$}&{$5.263\times 10^{-2}$}&{$5.657\times 10^{-2}$}&{$5.720\times 10^{-2}$}&{$5.734\times 10^{-2}$}&{$5.738\times 10^{-2}$}&{$5.739\times 10^{-2}$} \\ \hline
{$3$}&{$-$}&{$-$}&{$4.335\times 10^{-3}$}&{$4.820\times 10^{-3}$}&{$4.920\times 10^{-3}$}&{$4.946\times 10^{-3}$}&{$4.954\times 10^{-3}$} \\ \hline
{$4$}&{$-$}&{$-$}&{$-$}&{$3.671\times 10^{-4}$}&{$4.188\times 10^{-4}$}&{$4.313\times 10^{-4}$}&{$4.349\times 10^{-4}$} \\ \hline
{$5$}&{$-$}&{$-$}&{$-$}&{$-$}&{$2.966\times 10^{-5}$}&{$3.452\times 10^{-5}$}&{$3.583\times 10^{-5}$} \\ \hline
{$6$}&{$-$}&{$-$}&{$-$}&{$-$}&{$-$}&{$2.241\times 10^{-6}$}&{$2.651\times 10^{-6}$} \\ \hline
{$7$}&{$-$}&{$-$}&{$-$}&{$-$}&{$-$}&{$-$}&{$1.576\times 10^{-7}$} \\ \hline 
\end{tabular}
\end{center}
\end{table}

\newpage

\begin{table}[htbp]
\begin{center}
\caption{\label{b0r}Numerical constants $b_{0r}$ in
 eq.(\ref{be42AB})}
\begin{tabular}{@{\hspace{\tabcolsep}\extracolsep{\fill}}ccccccc} \hline 
{$r$}&{$r \le 2$}&{$r \le 3$}&{$r \le 4$}&{$r \le 5$}&{$r \le 6$} \\ \hline
{$2$}&{$2.778$}&{$2.530$}&{$2.497$}&{$2.490$}&{$2.488$} \\ \hline
{$3$}&{$-$}&{$-3.292\times 10^{-1}$}&{$-3.602\times 10^{-1}$}&{$-3.663\times 10^{-1}$}&{$-3.679\times 10^{-1}$} \\ \hline
{$4$}&{$-$}&{$-$}&{$-2.676 \times 10^{-2}$}&{$-3.038\times 10^{-2}$}&{$-3.125\times 10^{-2}$} \\ \hline
{$5$}&{$-$}&{$-$}&{$-$}&{$-2.294\times 10^{-3}$}&{$-2.670\times 10^{-3}$} \\ \hline
{$6$}&{$-$}&{$-$}&{$-$}&{$-$}&{$-1.882\times 10^{-4}$} \\ \hline 
\end{tabular}
\end{center}
\end{table}

\newpage

\begin{table}[htbp]
\begin{center}
\caption{\label{b2rAB}Numerical constants $b_{2r}^{A}$(upper) and $b_{2r}^{B}$(lower) in
 eq.(\ref{be45AB})}
\begin{tabular}{@{\hspace{\tabcolsep}\extracolsep{\fill}}cccccc} \hline 
{$r$}&{$r \le 2$}&{$r \le 3$}&{$r \le 4$}&{$r \le 5$}&{$r \le 6$}  \\ \hline
{$0$}&{$-1.277\times 10^{-2}$}&{$-1.784\times 10^{-3}$}&{$2.073\times 10^{-4}$}&{$6.982\times 10^{-4}$}&{$8.380\times 10^{-4}$} \\ \hline
{$1$}&{$-1.007$}&{$-9.775\times 10^{-1}$}&{$-9.721\times 10^{-1}$}&{$-9.708\times 10^{-1}$}&{$-9.704\times 10^{-1}$} \\ \hline
{$2$}&{$3.363\times 10^{-1}$}&{$3.701\times 10^{-1}$}&{$3.758\times 10^{-1}$}&{$3.772\times 10^{-1}$}&{$3.775\times 10^{-1}$} \\ \hline
{$3$}&{$-$}&{$3.262\times 10^{-2}$}&{$3.643\times 10^{-2}$}&{$3.729\times 10^{-2}$}&{$3.753\times 10^{-2}$} \\ \hline
{$4$}&{$-$}&{$-$}&{$2.584\times 10^{-3}$}&{$2.986\times 10^{-3}$}&{$3.089\times 10^{-3}$} \\ \hline                  
{$5$}&{$-$}&{$-$}&{$-$}&{$2.091\times 10^{-4}$}&{$2.461\times 10^{-4}$} \\ \hline
{$6$}&{$-$}&{$-$}&{$-$}&{$-$}&{$1.557\times 10^{-5}$} \\ \hline 
\end{tabular}
\end{center}
\begin{center}
\begin{tabular}{@{\hspace{\tabcolsep}\extracolsep{\fill}}cccccc} \hline 
{$r$}&{$r \le 2$}&{$r \le 3$}&{$r \le 4$}&{$r \le 5$}&{$r \le 6$}  \\ \hline
{$0$}&{$4.052\times 10^{-1}$}&{$4.041\times 10^{-1}$}&{$4.039\times 10^{-1}$}&{$4.038\times 10^{-1}$}&{$4.038\times 10^{-1}$} \\ \hline
{$1$}&{$-4.261\times 10^{-1}$}&{$-4.297\times 10^{-1}$}&{$-4.304\times 10^{-1}$}&{$-4.306\times 10^{-1}$}&{$-4.306\times 10^{-1}$} \\ \hline
{$2$}&{$-4.452\times 10^{-2}$}&{$-4.886\times 10^{-2}$}&{$-4.968\times 10^{-2}$}&{$-4.987\times 10^{-2}$}&{$-4.993\times 10^{-2}$} \\ \hline
{$3$}&{$-$}&{$-4.594\times 10^{-3}$}&{$-5.174\times 10^{-3}$}&{$-5.304\times 10^{-3}$}&{$-5.340\times 10^{-3}$} \\ \hline
{$4$}&{$-$}&{$-$}&{$-4.315\times 10^{-4}$}&{$-4.961\times 10^{-4}$}&{$-5.125 \times 10^{-4}$} \\ \hline                  
{$5$}&{$-$}&{$-$}&{$-$}&{$-3.688\times 10^{-5}$}&{$-4.311\times 10^{-5}$} \\ \hline
{$6$}&{$-$}&{$-$}&{$-$}&{$-$}&{$-2.881\times 10^{-6}$} \\ \hline 
\end{tabular}
\end{center}
\end{table}

\newpage

\begin{table}[htbp]
\begin{center}
\caption{\label{b2r}Numerical constants $b_{2r}$ in
 eq.(\ref{be45})}
\begin{tabular}{@{\hspace{\tabcolsep}\extracolsep{\fill}}cccccc} \hline 
{$r$}&{$r \le 2$}&{$r \le 3$}&{$r \le 4$}&{$r \le 5$}&{$r \le 6$}  \\ \hline
{$0$}&{$-2.154\times 10^{-1}$}&{$-2.038\times 10^{-1}$}&{$-2.017\times 10^{-1}$}&{$-2.012\times 10^{-1}$}&{$-2.011\times 10^{-1}$} \\ \hline
{$1$}&{$-7.943\times 10^{-1}$}&{$-7.626\times 10^{-1}$}&{$-7.569\times 10^{-1}$}&{$-7.555\times 10^{-1}$}&{$-7.551\times 10^{-1}$}\\ \hline
{$2$}&{$3.586\times 10^{-1}$}&{$3.945\times 10^{-1}$}&{$4.006\times 10^{-1}$}&{$4.021\times 10^{-1}$}&{$4.025\times 10^{-1}$} \\ \hline
{$3$}&{$-$}&{$3.492\times 10^{-2}$}&{$3.902\times 10^{-2}$}&{$3.994\times 10^{-2}$}&{$4.020\times 10^{-2}$} \\ \hline
{$4$}&{$-$}&{$-$}&{$2.800\times 10^{-3}$}&{$3.234\times 10^{-3}$}&{$3.346\times 10^{-3}$}\\ \hline
{$5$}&{$-$}&{$-$}&{$-$}&{$2.276\times 10^{-4}$}&{$2.677\times 10^{-4}$} \\ \hline
{$6$}&{$-$}&{$-$}&{$-$}&{$-$}&{$1.701\times 10^{-5}$} \\ \hline 
\end{tabular}
\end{center}
\end{table}

\newpage

\begin{table}[htbp]
\begin{center}
\caption{\label{macro1}Numerical constants for the macroscopic
 quantities: the $i$th approximation quantities for hard-disk molecules and the exact values for the steady-state BGK equation and information theory. }
\begin{tabular}{@{\hspace{\tabcolsep}\extracolsep{\fill}}cccccc} \hline 
{$i$th}&{$\lambda_{P}^{xx}$}&{$\lambda_{S}$}\\ \hline
{$3$th}&{$-5.085\times 10^{-2}$}&{$-2.549\times 10^{-1}$} \\ \hline
{$4$th}&{$-4.807\times 10^{-2}$}&{$-2.551\times 10^{-1}$} \\ \hline
{$5$th}&{$-4.757\times 10^{-2}$}&{$-2.551\times 10^{-1}$} \\ \hline
{$6$th}&{$-4.744\times 10^{-2}$}&{$-2.552\times 10^{-1}$} \\ \hline
{$7$th}&{$-4.741\times 10^{-2}$}&{$-2.552\times 10^{-1}$} \\ \hline
{BGK equation}&{$0$}&{$-\frac{1}{4}$}\\ \hline 
{information}&{$\frac{1}{2}$}&{$-\frac{1}{4}$} \\ \hline
\end{tabular}
\end{center}
\end{table}

\newpage

\begin{table}[htbp]
\begin{center}
\caption{\label{alpha}Numerical constants $\alpha_{r}$ in
 eq.(\ref{cr120}). }
\begin{tabular}{@{\hspace{\tabcolsep}\extracolsep{\fill}}ccc} \hline 
{$r$}&{Boltzmann Eq.}&{BGK Eq. and Information Theory} \\ \hline
{$0$}&{$-2.758 \times 10^{-2}$}&{$-\frac{3}{128}$} \\ \hline
{$1$}&{$-1.761\times 10^{-1}$}&{$-\frac{9}{64}$} \\ \hline
{$2$}&{$4.015\times 10^{-1}$}&{$\frac{9}{32}$} \\ \hline
{$3$}&{$-1.310\times 10^{-1}$}&{$-\frac{1}{16}$} \\ \hline
{$4$}&{$1.326\times 10^{-2}$}&{$-$} \\ \hline
{$5$}&{$-1.368\times 10^{-3}$}&{$-$} \\ \hline
{$6$}&{$1.260\times 10^{-4}$}&{$-$} \\ \hline
{$7$}&{$-9.809\times 10^{-6}$}&{$-$} \\ \hline
{$8$}&{$6.290\times 10^{-7}$}&{$-$} \\ \hline
{$9$}&{$-3.249\times 10^{-8}$}&{$-$} \\ \hline
{$10$}&{$1.306\times 10^{-9}$}&{$-$} \\ \hline
{$11$}&{$-3.902\times 10^{-11}$}&{$-$} \\ \hline
{$12$}&{$8.256\times 10^{-13}$}&{$-$} \\ \hline
{$13$}&{$-1.160\times 10^{-14}$}&{$-$} \\ \hline
{$14$}&{$9.645\times 10^{-17}$}&{$-$} \\ \hline
{$15$}&{$-3.576\times 10^{-19}$}&{$-$} \\ \hline
\end{tabular}
\end{center}
\end{table}

\newpage

\begin{table}[htbp]
\begin{center}
\caption{\label{beta}Numerical constants $\beta_{r}$ in eq.(\ref{cr130}).}
\begin{tabular}{@{\hspace{\tabcolsep}\extracolsep{\fill}}cccc} \hline 
{$r$}&{Boltzmann Eq.}&{BGK Eq.}&{Information Theory} \\ \hline
{$0$}&{$-1.284\times 10^{-1}$}&{$-\frac{9}{64}$}&{$-\frac{17}{128}$} \\ \hline
{$1$}&{$-4.721\times 10^{-1}$}&{$-\frac{30}{64}$}&{$-\frac{31}{64}$} \\ \hline
{$2$}&{$3.372\times 10^{-1}$}&{$\frac{3}{16}$}&{$\frac{11}{32}$} \\ \hline
{$3$}&{$7.086\times 10^{-2}$}&{$\frac{1}{8}$}&{$\frac{1}{16}$} \\ \hline
{$4$}&{$-3.874\times 10^{-3}$}&{$-$}&{$-$} \\ \hline
{$5$}&{$1.536\times 10^{-4}$}&{$-$}&{$-$} \\ \hline
{$6$}&{$-2.774\times 10^{-6}$}&{$-$}&{$-$} \\ \hline
\end{tabular}
\end{center}
\end{table}

\end{document}